\documentclass[floatfix,nofootinbib,superscriptaddress,aps,prc]{revtex4-1} 
\usepackage[utf8]{inputenc}
\usepackage{multido}
\usepackage{mathrsfs}
\usepackage{amsmath}
\usepackage{amssymb}
\usepackage{amsfonts}
\usepackage{xspace}
\usepackage{graphicx}		
\usepackage{graphics}
\usepackage{units}
\usepackage{hyperref}
\usepackage{ulem}

\newcommand{\vect}[1]{\vec{\boldsymbol{#1}}}

\newcommand{\comment}[1]{}

\newcommand{\EFT}{$\mathrm{EFT}_{\not{\pi}}$\xspace}
\newcommand{\SixJ}[6]{\left\{\begin{array}{ccc} #1 & #2 & #3\\[1 mm] #4 & #5 & #6 \end{array}\right\}}
\newcommand{\CG}[6]{C_{#1,#3,#5}^{#2,#4,#6}}

\newcommand{\wave}[3]{\ensuremath{{}^{#1}\mathrm{#2}_{#3}}\xspace}

\newcommand{\oneP}{\wave{1}{P}{1}}
\newcommand{\threePzero}{\wave{3}{P}{0}}
\newcommand{\threePone}{\wave{3}{P}{1}}
\newcommand{\threePtwo}{\wave{3}{P}{2}}

\newcommand{\jjvHe}{{}^3\mathrm{He}}

\newcommand{\Pc}{\hat{\mathcal{P}}}

\renewcommand{\emph}[1]{\textit{#1}}
\newcommand{\N}[1]{\mathrm{N}^{#1}\mathrm{LO}}

\newcommand{\nnlo}{\mathrm{N}^{2}\mathrm{LO}}
\newcommand{\ntlo}{\mathrm{N}^{3}\mathrm{LO}}
\newcommand{\tso}{{}^{3}\!S_{1}}
\newcommand{\osz}{{}^{1}\!S_{0}}

\newcommand{\SJ}[6]{\left\{\begin{array}{ccc} #1 & #2 & #3\\[1mm] #4 & #5 & #6 \end{array}\right\}}
\newcommand{\NJ}[9]{\left\{\begin{array}{ccc} #1 & #2 & #3\\[1mm] #4 & #5 & #6 \\[1mm] #7 & #8 & #9 \end{array}\right\}}

\begin{document}

\title{$nd$ Scattering and the $A_y$ Puzzle to Next-to-next-to-next-to-leading Order}

\author{Arman Margaryan}
\email{am343@phy.duke.edu}
\affiliation{Department of Physics, Duke University, Durham, NC 27708, USA}

\author{Roxanne P. Springer}
\email{rps@phy.duke.edu}
\affiliation{Department of Physics, Duke University, Durham, NC 27708, USA}

\author{Jared Vanasse}
\email{jjv9@phy.duke.edu}
\email{vanasse@ohio.edu}
\affiliation{Department of Physics, Duke University, Durham, NC 27708, USA}
\affiliation{Department of Physics and Astronomy, Ohio University, Athens, OH 45701, USA}

\date{\today}

\begin{abstract}
Polarization observables in neutron-deuteron scattering are calculated to next-to-next-to-next-to-leading order ($\ntlo$) in pionless effective field theory (\EFT).  At $\ntlo$ the two-body $P$-wave contact interactions are found to be important contributions to the neutron vector analyzing power, $A_{y}(\theta)$, and the deuteron vector analyzing power, $iT_{11}(\theta)$.  Extracting the two-body $P$-wave \EFT coefficients from two-body scattering data and varying them within the expected \EFT theoretical errors provides results that are consistent (at the $\ntlo$ level) with $A_y$ experimental data at low energies.  Cutoff dependence of the $\ntlo$ correction of the doublet $S$-wave $nd$ scattering amplitude suggests the need for a new three-body force at $\ntlo$, which is likely one that mixes Wigner-symmetric and Wigner-antisymmetric three-body channels.
\end{abstract}

\keywords{latex-community, revtex4, aps, papers}

\maketitle

\section{Introduction}

The maximum of the nucleon vector analyzing power, $A_{y}(\theta)$, and the deuteron vector analyzing power, $iT_{11}(\theta)$, in neutron-deuteron ($nd$) and proton-deuteron ($pd$) scattering are significantly underpredicted by existing three-body calculations at low energies (nucleon laboratory energies below $30$~MeV).  This is known as the three-nucleon analyzing power problem or as the $A_{y}$ puzzle (See, e.g., Refs.~\cite{Entem:2001tj,Kievsky:2013yva}). Both phenomenological potential model calculations (PMC)~\cite{Kievsky:2013yva} and more modern potentials derived from chiral effective field theory~\cite{Binder:2015mbz} have not resolved this discrepancy.  PMC have shown that $A_{y}$ is very sensitive to the values of the two-body ${}^{3}\!P_{J}$ phase shifts~\cite{Witala1989446,WITALA199148,TORNOW1991273} and that the experimental neutron-proton ($np$) scattering data does not give enough latitude to simultaneously fit the two-body ${}^{3}\!P_{J}$ wave phase shifts and $A_{y}$~\cite{Huber:1998hu}. 

For low energies ($E<m_{\pi}^{2}/M_{N}$) nuclear systems can be described by a theory containing only contact interactions between nucleons and possible external currents.  This theory, known as pionless effective field theory (\EFT) (See, e.g. Ref.~\cite{Beane:2000fx} for a review), has been used to calculate nucleon-nucleon ($NN$) scattering~\cite{Chen:1999tn,Kong:1999sf}, deuteron electromagnetic form factors~\cite{Chen:1999tn}, $np$ capture~\cite{Rupak:1999rk}, and neutrino-deuteron scattering in the two-body sector~\cite{Butler:2000zp}.  Progress in the three-body sector includes the calculation of $nd$ scattering~\cite{Gabbiani:1999yv,Griesshammer:2004pe,Vanasse:2013sda} and $pd$ scattering \cite{Rupak:2001ci,Konig:2011yq,Vanasse:2014kxa,Konig:2014ufa}.  The first calculations of $nd$ scattering relied on the partial resummation technique~\cite{Bedaque:2002yg}, which resummed certain higher order contributions and therefore were not strictly perturbative in the \EFT power counting.  References~\cite{Vanasse:2013sda,Vanasse:2015} introduced a technique to calculate $nd$ scattering amplitudes strictly perturbatively that is no more numerically expensive than the partial resummation technique.  With the strictly perturbative approach $nd$ scattering was calculated to next-to-next-to-leading order ($\nnlo$) including the two-body $SD$-mixing term~\cite{Vanasse:2013sda}.  The two-body $SD$-mixing term provides the first non-zero contribution to the polarization observables in $nd$ scattering.  However, that work did not investigate polarization observables since they are not expected to be reproduced well at $\nnlo$.  The $\nnlo$ calculation is LO in the polarization observables since it gives the first non-zero contribution to them.

Building on this $\nnlo$ calculation, we present in this paper results for polarization observables in $nd$ scattering to next-to-next-to-next-to-leading order ($\ntlo$) in \EFT.  At $\ntlo$ there are new contributions from shape parameter corrections as well as the two-body $P$-wave contact interactions that are found to give the dominant contribution to $A_{y}$ (as already identified in PMC).  At $\nnlo$ a term that causes splitting between the neutron-neutron ($nn$) and $np$ $\osz$ scattering lengths should be included.  Experimentally the difference in scattering lengths is roughly 3\% and therefore numerically is considered a $\nnlo$ correction in the \EFT power counting.  This term has been included in previous two-body calculations~\cite{Rupak:1999rk}, but was not treated strictly perturbatively as it is in this work.

This paper is organized as follows.  In Sec.~\ref{sec:2B} all necessary two-body physics for the three-body calculation is presented. Section~\ref{sec:3B} discusses how the $nd$ scattering amplitudes are calculated and introduces the concept of $P$-wave auxiliary fields to calculate the three-body contribution from the two-body $P$-wave contact interactions.  In Sec.~\ref{sec:observables} expressions for the polarization observables in $nd$ scattering are introduced.  Polarization observable results in \EFT are shown in Sec.~\ref{sec:results}, and finally we conclude in Sec.~\ref{sec:conclusion}.

\section{\label{sec:2B}Two-Body Scattering}
The two-body $S$-wave Lagrangian up to and including $\ntlo$ in the $Z$-parametrization~\cite{Phillips:1999hh,Griesshammer:2004pe} is given by
\begin{align}
\label{eq:lagS}
\mathcal{L}_{2}^{S}=\ &\hat{N}^{\dagger}\left(i\partial_{0}+\frac{\vect{\nabla}^2}{2M_{N}}\right)\hat{N}\\\nonumber
&+\hat{t}_{i}^{\dagger}\left(\Delta_{t}-c_{0t}\left(i\partial_{0}+\frac{\vect{\nabla}^{2}}{4M_{N}}+\frac{\gamma_{t}^{2}}{M_{N}}\right)-c_{1t}\left(i\partial_{0}+\frac{\vect{\nabla}^{2}}{4M_{N}}+\frac{\gamma_{t}^{2}}{M_{N}}\right)^{2}\right)\hat{t}_{i}\\\nonumber
&+\hat{s}_{a}^{\dagger}\left(\Delta_{s}+\Delta_{s}^{(\nnlo)}\delta_{-1}^{a}-c_{0s}\left(i\partial_{0}+\frac{\vect{\nabla}^{2}}{4M_{N}}+\frac{\gamma_{s}^{2}}{M_{N}}\right)-c_{1s}\left(i\partial_{0}+\frac{\vect{\nabla}^{2}}{4M_{N}}+\frac{\gamma_{s}^{2}}{M_{N}}\right)^{2}\right)\hat{s}_{a}\\\nonumber
&+y_{t}\left[\hat{t}_{i}^{\dagger}\hat{N} ^{T}P_{i}\hat{N} +\mathrm{H.c.}\right]+y_{s}\left[\hat{s}_{a}^{\dagger}\hat{N}^{T}\bar{P}_{a}\hat{N}+\mathrm{H.c.}\right],
\end{align}
where $\hat{t}_{i}$ ($\hat{s}_{a}$) is the spin-triplet iso-singlet (spin-singlet iso-triplet) auxiliary field, and $P_{i}=\frac{1}{\sqrt{8}}\sigma_{2}\sigma_{i}\tau_{2}$ ($\bar{P}_{a}=\frac{1}{\sqrt{8}}\sigma_{2}\tau_{2}\tau_{a}$) projects out the spin-triplet iso-singlet (spin-singlet iso-triplet) combination of nucleons.  The subscript ``2'' indicates that Eq.~(\ref{eq:lagS}) includes only two-body terms.  At LO the parameters are fit to reproduce the bound state and virtual bound state poles in the $\tso$ and $\osz$ channels, respectively.  The majority of NLO, $\nnlo$, and $\ntlo$ parameters in the $Z$-parametrization are then fit to ensure the poles remain unchanged and have the correct residues.  The $\ntlo$ parameters $c_{1t}$ and $c_{1s}$ are fit to reproduce the effective range expansion (ERE) shape corrections about the poles of the $\tso$ and $\osz$ channels, respectively.  Carrying out this procedure yields~\cite{Griesshammer:2004pe}
\begin{align}
&\Delta_{t}+\mu=\gamma_{t},\,\,\, y_{t}=\sqrt{\frac{4\pi}{M_{N}}},\,\,\, c_{0t}^{(n)}=(-1)^{n}(Z_{t}-1)^{n+1}\frac{M_{N}}{2\gamma_{t}},\,\,\, c_{1t}=\rho_{1t}M_{N}^{2}\\\nonumber
&\Delta_{s}+\mu=\gamma_{s},\,\,\, y_{s}=\sqrt{\frac{4\pi}{M_{N}}},\,\,\, c_{0s}^{(n)}=(-1)^{n}(Z_{s}-1)^{n+1}\frac{M_{N}}{2\gamma_{s}},\,\,\, c_{1s}=\rho_{1s}M_{N}^{2},
\end{align}
where $\mu$ is a scale introduced via dimensional regularization with the power divergence subtraction scheme~\cite{Kaplan:1998tg,Kaplan:1998we}.  All physical observables must be $\mu$-independent.  The value $\gamma_{t}=45.7025$~MeV ($\gamma_{s}=-7.890$~MeV) is the deuteron bound state momentum ($\osz$ virtual bound state momentum), and $Z_{t}=1.6908$ ($Z_{s}=0.9015$) the residue of the $\tso$ ($\osz$) bound state (virtual bound state) pole.  For the shape parameter correction about the $\tso$ ($\osz$) pole we use $\rho_{1t}=0.389~\mathrm{fm}^{3}$ ($\rho_{1s}=-0.48~\mathrm{fm}^{3}$).  The $\nnlo$ parameter $\Delta_{s}^{(\nnlo)}=-2.02$~MeV, and is fit to the splitting between the virtual bound state momentum in the $np$ and $nn$ spin-singlet channels.  There is also a separate parameter for the splitting between the virtual bound state momentum in the $np$ and proton-proton ($pp$) spin-singlet channels in the absence of Coulomb, but this is not relevant for the $nd$ system and is not considered here.

At $\nnlo$ there is a  contribution from two-body $SD$-mixing given by the Lagrangian
\begin{equation}
\mathcal{L}^{SD}_{2}=y_{SD}\hat{t}_{i}^{\dagger}\left[\hat{N}^{T}\left((\stackrel{\rightarrow}{\partial}-\stackrel{\leftarrow}{\partial})^{i}(\stackrel{\rightarrow}{\partial}-\stackrel{\leftarrow}{\partial})^{j}-\frac{1}{3}\delta^{ij}(\stackrel{\rightarrow}{\partial}-\stackrel{\leftarrow}{\partial})^{2}\right)P_{j}\hat{N}\right]+\mathrm{H.c.}.
\end{equation}
The parameter $y_{SD}$ is fit to the asymptotic $D/S$ mixing ratio of the deuteron wavefunction, yielding~\cite{Vanasse:2013sda,Chen:1999tn}
\begin{equation}
y_{SD}=-\sqrt{\frac{4\pi}{M_{N}}}\frac{3\eta_{sd}\sqrt{2}}{8\gamma_{t}^{2}},
\end{equation}
where $\eta_{sd}=.02543\pm.00007$ is the asymptotic $D/S$ mixing ratio of the deuteron wavefunction~\cite{Stoks:1994wp}.  

Two-body $P$-wave contact interactions first occur at $\ntlo$.  The ${}^{3}\!P_{J}$ terms are given by the ${}^{3}\!P_{J}$ Lagrangian~\cite{Chen:1999bg},  
\begin{align}
\label{eq:PwaveLagRupak}
&\mathcal{L}_{2}^{{}^{3}P_{J}}=\left(C_{2}^{({}^3\!P_{0})}\delta_{xy}\delta_{wz}+C_{2}^{({}^3\!P_{1})}[\delta_{xw}\delta_{yz}-\delta_{xz}\delta_{yw}]+C_{2}^{({}^3\!P_{2})}\left[2\delta_{xw}\delta_{yz}+2\delta_{xz}\delta_{yw}-\frac{4}{3}\delta_{xy}\delta_{wz}\right]\right)\\\nonumber
&\hspace{3cm}\times\frac{1}{4}(\hat{N}^{T}\mathcal{O}_{xyA}^{(1,P)}\hat{N})^{\dagger}(\hat{N}^{T}\mathcal{O}_{wzA}^{(1,P)}\hat{N}),
\end{align}
where
\begin{equation}
\mathcal{O}_{ijA}^{(1,P)}=\stackrel{\leftarrow}{\nabla}_{i}P_{jA}^{P}-P_{jA}^{P}\stackrel{\rightarrow}{\nabla}_{i}
\end{equation}
and the projector is defined as $P_{iA}^{P}=\frac{1}{\sqrt{8}}\sigma_{2}\sigma_{i}\tau_{2}\tau_{A}$.  Note that the projector $P_{iA}^{P}$ differs from the projector in Ref.~\cite{Chen:1999bg} because we consider $NN$ scattering, not just $np$ scattering.  At $\ntlo$ the two-body ${}^{1}\!P_{1}$  contact interaction also appears:
\begin{equation}
\label{eq:Pwave1P1}
 \mathcal{L}_{2}^{{}^{1}\!P_{1}}=C_{2}^{({}^{1}\!P_{1})}\frac{1}{4}(\hat{N}^{T}\mathcal{O}_{x}^{(0,P)}\hat{N})^{\dagger}(\hat{N}^{T}\mathcal{O}_{x}^{(0,P)}\hat{N}),
\end{equation}
but it does not contribute to the polarization observables in our calculation at this order.  The operator $\mathcal{O}_{i}^{(0,P)}$ is defined by
\begin{equation}
\mathcal{O}_{i}^{(0,P)}=\stackrel{\leftarrow}{\nabla}_{i}P^{P}-P^{P}\stackrel{\rightarrow}{\nabla}_{i},
\end{equation}
where the projector is $P^{P}=\frac{1}{\sqrt{8}}\sigma_{2}\tau_{2}$.  Fitting the coefficients to the $np$ Nijmegen phase shifts~\cite{Stoks:1994wp} yields the values 
\begin{equation}
\label{eq:CPJnpfit}
C^{{}^{3}\!P_{0}}=6.27~\mathrm{fm}^{4}, \ \ C^{{}^{3}\!P_{1}}=-5.75~\mathrm{fm}^{4}, \ \ C^{{}^{3}\!P_{2}}=0.522~\mathrm{fm}^{4},\,\,\mathrm{and}\,\, C^{{}^{1}\!P_{1}}=-19.8~\mathrm{fm}^{4}.  
\end{equation}
These $C^{{}^{3}\!P_{J}}$ values are in good agreement with those found in Ref.~\cite{Chen:1999bg}.  We take them as  the central values of an experimental fit, but there is a substantial theoretical \EFT error associated with that fit; these coefficients are $\ntlo$ for $nd$ scattering but are LO in two-body $P$-wave scattering.

At LO the power counting mandates that an infinite number of diagrams be summed, yielding the LO dibaryon propagators~\cite{Vanasse:2013sda,Griesshammer:2004pe}
\begin{equation}
iD_{\{t,s\}}(p_{0},\vect{p})=\frac{i}{\gamma_{\{t,s\}}-\sqrt{\frac{\vect{p}^{2}}{4}-M_{N}p_{0}-i\epsilon}}.
\end{equation}
The LO deuteron wavefunction renormalization is the residue about the $\tso$ bound state pole, which gives
\begin{equation}
Z_{\mathrm{LO}}=\frac{2\gamma_{t}}{M_{N}}.
\end{equation}
The form of higher order dibaryon propagators and wavefunction renormalization constants can be found in Refs.~\cite{Vanasse:2013sda,Griesshammer:2004pe}.  For this work the form of these higher order corrections will not be explicitly needed.  Rather, the higher order corrections will naturally be included in the integral equations, and diagrams with corrections attached to external dibaryon propagators will give higher order deuteron wavefunction renormalization contributions in the on-shell limit.


\section{\label{sec:3B} Three-Body Scattering}

\begin{figure}[hbt]
\begin{center}
\includegraphics[width=100mm]{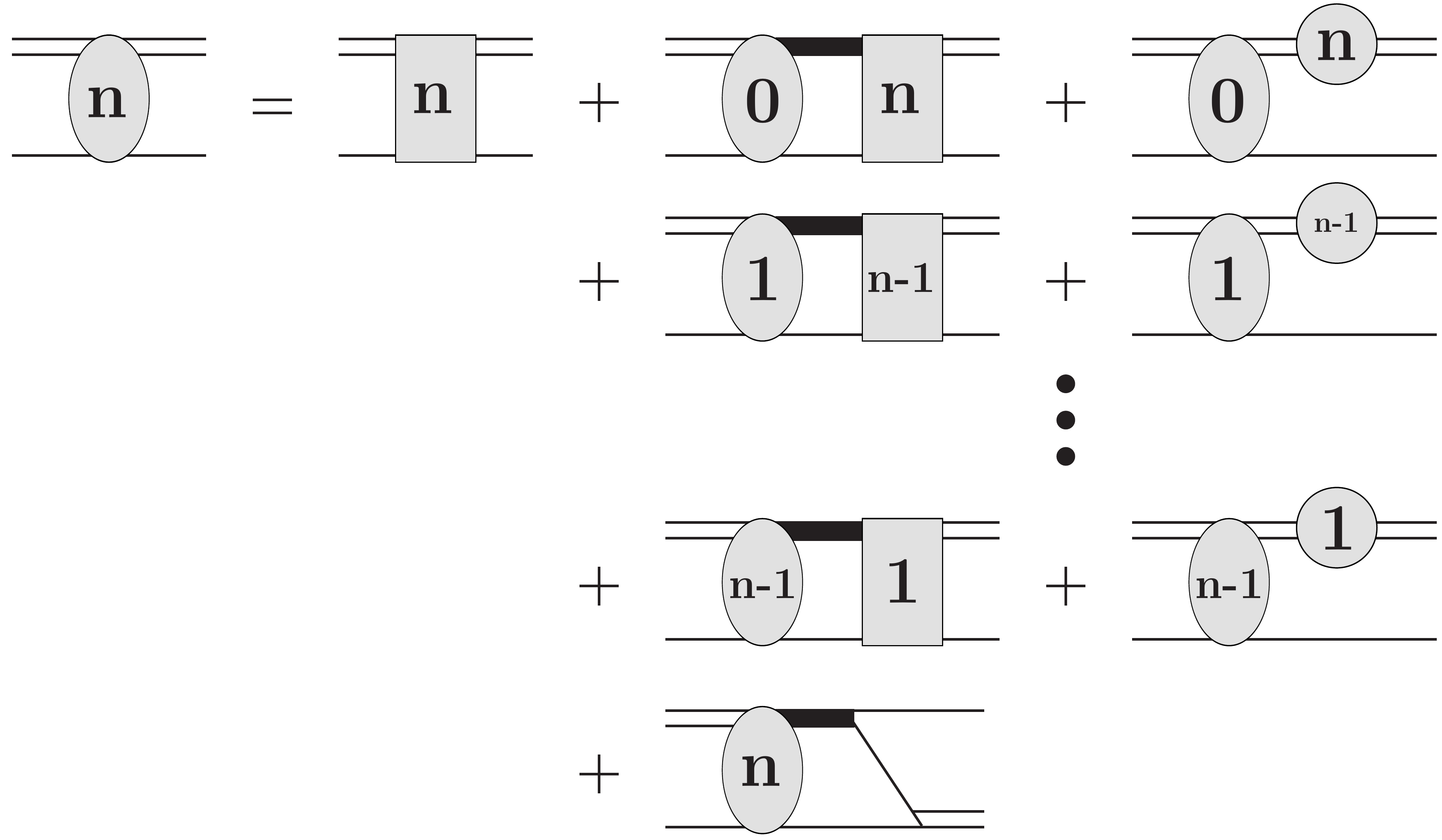}
\includegraphics[width=100mm]{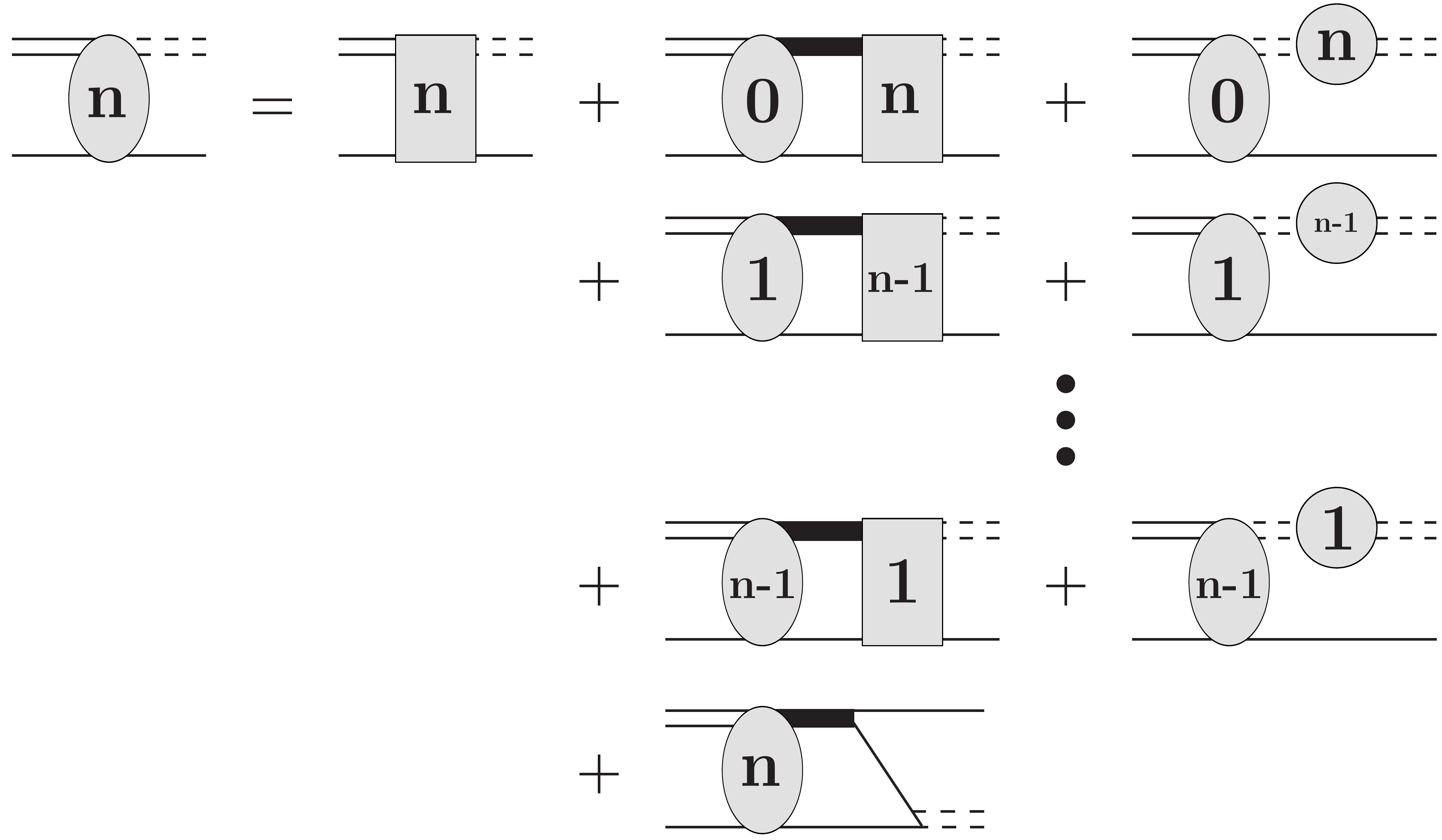}
\end{center}
\caption{\label{fig:PertCorrectionDiagrams}Single lines represent nucleons, double lines spin-triplet dibaryons, and double dashed lines spin-singlet dibaryons.  Thick solid lines denote a sum over both spin-triplet and spin-singlet dibaryons.  The LO $nd$ scattering amplitude is the oval with a ``0" inside, and the oval with the $n$ inside is the $\mathrm{N}^{n}\mathrm{LO}$ correction to the $nd$ scattering amplitude.  The circle with the $n$ inside is the $\mathrm{N}^{n}\mathrm{LO}$ correction to the dibaryon propagators (see Fig.~\ref{fig:TwoBodyContributions}), and the rectangle with the $n$ inside is the $n$th order ``three-body" correction (see Fig.~\ref{fig:N3LOndScatteringDiagrams}).}
\end{figure}

The $\mathrm{N}^{n}\mathrm{LO}$ correction to the $nd$ scattering amplitude using the methods introduced in Ref.~\cite{Vanasse:2015} is given by the integral equation represented in Fig.~\ref{fig:PertCorrectionDiagrams}.  Because this integral equation is not yet projected in spin or partial waves, it includes both doublet and quartet channel contributions.  The single line represents a nucleon, the double line a spin-triplet dibaryon, and the double dashed line a spin-singlet dibaryon.  A thick solid line denotes a sum over both spin-triplet and spin-singlet dibaryons.  The oval with a ``0" inside is the LO $nd$ scattering amplitude, the oval with the $n$ inside is the $n$th order correction to the $nd$ scattering amplitude, the circle with the $n$ inside is a $\mathrm{N}^{n}\mathrm{LO}$ correction to the dibaryon propagator, and the rectangle with the $n$ inside is a $\mathrm{N}^{n}\mathrm{LO}$ ``three-body" correction.\footnote{The term ``three-body" refers to corrections that involve all three-nucleons; this includes three-body forces.}

\begin{figure}[hbt]
\includegraphics[width=110mm]{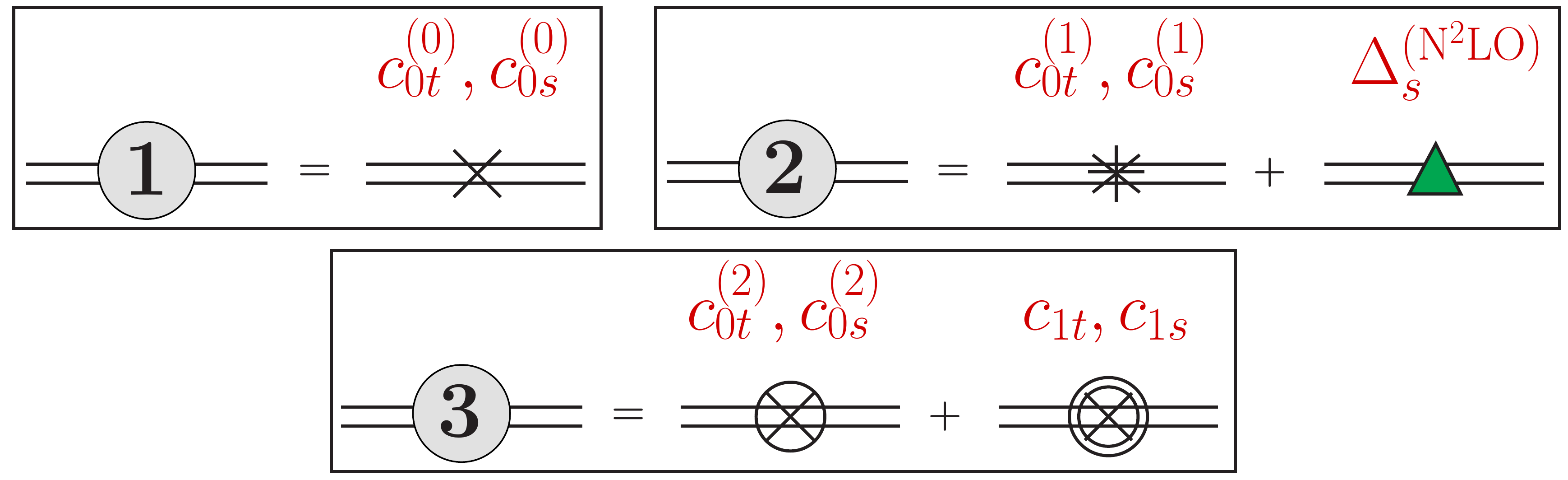}
\vspace{-.5cm}
\caption{\label{fig:TwoBodyContributions}Higher order corrections to dibaryon propagators (used in the diagrams of Fig.~\ref{fig:PertCorrectionDiagrams}).  The NLO (n=1) corrections are range corrections from $c_{0t}^{(0)}$ and $c_{0s}^{(0)}$.  At $\nnlo$ (n=2) the dibaryons receive further range corrections $c_{0t}^{(1)}$ and $c_{0s}^{(1)}$ in the $Z$-\\[.5mm]parametrization, as well as the $\Delta^{(\nnlo)}$ correction from splitting between the $nn$ and $np$ spin-singlet scattering lengths.  The $\ntlo$ (n=3) corrections are from higher order range corrections $c_{0t}^{(2)}$ and $c_{0s}^{(2)}$ in the $Z$-parametrization, and shape parameter corrections $c_{1t}$ and $c_{1s}$.  }
\end{figure}

Perturbative corrections to the dibaryon propagators in the $Z$-parametrization~\cite{Griesshammer:2004pe} are given in Fig.~\ref{fig:TwoBodyContributions}.  
The NLO contributions $c_{0t}^{(0)}$ and $c_{0s}^{(0)}$ are from range corrections, and the $\nnlo$ terms $c_{0t}^{(1)}$ and $c_{0s}^{(1)}$ are from higher order corrections to $c_{0t}^{(0)}$ and $c_{0s}^{(0)}$, respectively.  The $\nnlo$ correction, $\Delta_{s}^{(\nnlo)}$, arises from the splitting between the ${}^{1}\!S_{0}$ scattering length for $nn$ and $np$ scattering.\footnote{In the ERE there is no $\nnlo$ correction to the dibaryon propagator from $c_{0t}^{(1)}$ and $c_{0s}^{(1)}$ in this formalism.  However, there is still a correction from the splitting term $\Delta_{s}^{(\nnlo)}$ in the ERE.}  At $\mathrm{N}^{3}\mathrm{LO}$ there are corrections $c_{0t}^{(2)}$ and $c_{0s}^{(2)}$ to the effective range corrections and also shape parameter corrections $c_{1t}$ and $c_{1s}$ . 

\begin{figure}[hbt]
\includegraphics[width=105mm]{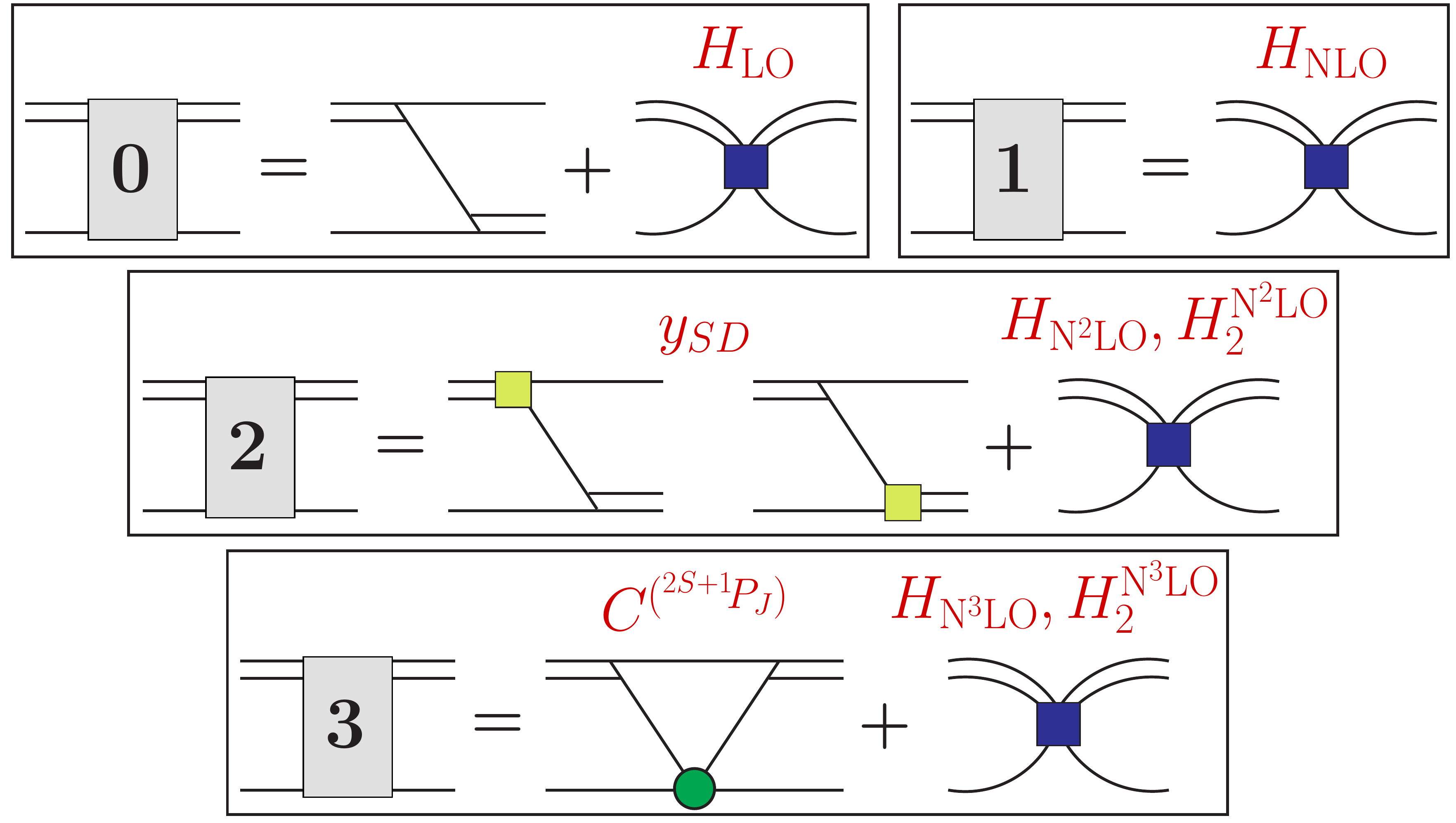}
\caption{\label{fig:N3LOndScatteringDiagrams}``three-body" contributions to integral equations (used in the diagrams of Fig.~\ref{fig:PertCorrectionDiagrams}).  The LO terms are nucleon exchange plus in the doublet $S$-wave channel the LO three-body force (dark square).  The NLO term is a NLO correction to the LO three-body force.  At $\nnlo$ there are contributions from the two-body $SD$-mixing term (coupling indicated by pale square), the $\nnlo$ correction to the LO three-body force, $H_{\nnlo}$, and a new energy dependent three-body force, $H_{2}^{\nnlo}$.  The $\ntlo$ contributions are from the two-body $P$-wave contact interactions (green circle), the $\ntlo$ correction to the LO three-body force, and the $\ntlo$ correction to the $\nnlo$ energy dependent three-body force.}
\end{figure}

The ``three-body" contributions to the integral equation are given by the diagrams in Fig.~\ref{fig:N3LOndScatteringDiagrams}.
LO diagrams are given by nucleon exchange and the LO three-body force in the doublet $S$-wave channel.  The NLO contribution comes from a NLO correction to the LO three-body force in the doublet $S$-wave channel.  At $\nnlo$ there is a contribution from the two-body $SD$-mixing term, a $\nnlo$ correction to the LO three-body force, and a new energy dependent three-body force in the the doublet $S$-wave channel.  Finally, at $\ntlo$ there are contributions from the $C^{{}^{3}\!P_{J}}$ and $C^{{}^{1}\!P_{1}}$ two-body $P$-wave contact interactions as well as $\ntlo$ corrections to the LO three-body force and the energy dependent $\nnlo$ three-body force.  Up to $\ntlo$ all three-body forces occur in the doublet $S$-wave channel. 

Projecting the $n$th order amplitude for $nd$ scattering in Fig.~\ref{fig:PertCorrectionDiagrams} into a partial wave basis yields the set of integral equations in cluster configuration (c.c.)~\cite{Griesshammer:2004pe} space 
\begin{align}
&\mathbf{t}_{n;L'S',LS}^{J}(k,p,E)=\mathbf{K}_{n;L'S',LS}^{J}(k,p,E)\mathbf{v}_{p}+\sum_{i=1}^{n}\mathbf{t}_{n-i;L'S',LS}^{J}(k,p,E)\circ\mathbf{R}_{i}(p,E)\\\nonumber
&+\sum_{L'',S''}\sum_{i=0}^{n-1}\mathbf{K}_{n-i;L'S',L''S''}^{J}(q,p,E)\  \mathbf{D} \!\left(\! E-\frac{q^{2}}{2M_{N}},\vect{q}\right)\otimes\mathbf{t}_{i;L''S'',LS}^{J}(k,q,E)\\\nonumber
&+\sum_{L'',S''}\mathbf{K}_{0;L'S',L''S''}^{J}(q,p,E) \ \mathbf{D} \! \left(\! E-\frac{q^{2}}{2M_{N}},\vect{q}\right)\otimes\mathbf{t}_{n;L''S'',LS}^{J}(k,q,E),
\end{align}
where
\begin{equation}
\label{eq:DibMatrix}
\mathbf{D}(E,\vect{q})=
\left(
\begin{array}{cc}
D_{t}(E,\vect{q}) & 0 \\
0&D_{s}(E,\vect{q})  
\end{array}\right)
\end{equation}
is a matrix of LO dibaryon propagators in c.c.~space, and the first subscript in all terms appearing in the integral equations refers to the order of a term ($n=0$ is LO, $n=1$ is NLO, etc.).  $\mathbf{t}^{J}_{n,L'S',LS}(k,p,E)$ and $\mathbf{v}_{p}$ are vectors in c.c.~space defined by
\begin{equation}
\hspace{-.85cm}\label{eq:tDef}
\mathbf{t}^{J}_{n,L'S',LS}(k,p,E)=\left(
\begin{array}{c}
t^{J;Nt\to Nt}_{n;L'S',LS}(k,p,E)\\[2mm]
t^{J;Nt\to Ns}_{n;L'S',LS}(k,p,E)
\end{array}\right), \ \ 
\mathbf{v}_{p}=\left(\begin{array}{c}
1 \\
0
\end{array}\right),
\end{equation}
where $t^{J;Nt\to Nt}_{n;L'S',LS}(k,p,E)$  is the amplitude for $nd$ scattering, $t^{J;Nt\to Ns}_{n;L'S',LS}(k,p,E)$ the amplitude for  a neutron and deuteron going to a nucleon and spin-singlet dibaryon, and $\mathbf{v}_{p}$ projects out diagrams in c.c.~space corresponding to a spin-triplet dibaryon in the initial state. The value $L$ ($S$) refers to the initial orbital (total spin) angular momentum, $L^\prime$ ($S^\prime$) to the final orbital (total spin) angular momentum, and $J$ to the total angular momentum (orbital plus total spin angular momentum).  All integral equations are calculated half off-shell in the center of mass (c.m.) frame, with $p$ being the outgoing off-shell momentum and $k$ the incoming on-shell momentum such that the total energy of the $nd$ system is given by $E=\frac{3}{4}\frac{k^{2}}{M_{N}}-\frac{\gamma_{t}^{2}}{M_{N}}$.  For this calculation all partial waves up to $L=4$ are included.  All three-body\\[.5mm] integrals are regulated using a sharp cutoff, which in the ``$\otimes$" notation is defined by
\begin{equation}
A(q)\otimes B(q)=\frac{1}{2\pi^{2}}\int_{0}^{\Lambda}dq \, q^{2}A(q)B(q).
\end{equation}
The ``$\circ$'' notation defines the Schur product (element wise matrix multiplication) of c.c.~space vectors.  For this set of integral equations the LO kernel in c.c.~space is given by
\begin{align}
&\mathbf{K}^{J}_{0;L'S',LS}(q,p,E)=\\\nonumber
&\hspace{1cm}\delta_{LL'}\delta_{SS'}\left\{\begin{array}{cc}
-\frac{2\pi}{qp}Q_{L}\left(\frac{q^{2}+p^{2}-M_{N}E-i\epsilon}{qp}\right)\left(
\begin{array}{rr}
1 & -3 \\
-3 & 1
\end{array}\right)-\pi H_{\mathrm{LO}}\delta_{L0}
\left(\!\!\begin{array}{rr}
1 & -1 \\
-1 & 1 
\end{array}\!\right) & ,\ \ S=\nicefrac{1}{2}\\[8mm]
-\frac{4\pi}{qp}Q_{L}\left(\frac{q^{2}+p^{2}-M_{N}E-i\epsilon}{qp}\right)\left(
\begin{array}{cc}
1 & 0\\
0 & 0
\end{array}\right) & ,\ \ S=\nicefrac{3}{2}\\
\end{array}\right.,
\end{align}
where at this order there is no mixing between different partial waves or splitting of different $J$-values.  The LO three-body force, $H_{\mathrm{LO}}$, is fit to the doublet $S$-wave $nd$ scattering length, $a_{nd}=0.65$~fm.  For details of how the fit is performed see Ref.~\cite{Vanasse:2015}.  The functions $Q_{L}(a)$ are Legendre functions of the second kind and are defined as\footnote{The definition of $Q_{L}(a)$ used here differs from the conventional definition by a phase factor of $(-1)^{L}$.}
\begin{equation}
Q_{L}(a)=\frac{1}{2}\int_{-1}^{1}dx\frac{P_{L}(x)}{x+a}, 
\end{equation}
with $P_{L}(x)$ being the standard Legendre polynomials.  Note that $\mathbf{R}_{0}(p,E)$ does not exist.

At NLO the only contribution to the kernel comes from the NLO correction to the three-body force, yielding
\begin{equation}
\hspace{-.5cm}\mathbf{K}^{J}_{1;L'S',LS}(q,p,E)=-\pi H_{\mathrm{NLO}}\delta_{L0}\delta_{LL'}\delta_{SS'}\delta_{S\nicefrac{1}{2}}
\left(\!\!\begin{array}{rr}
1 & -1 \\
-1 & 1 
\end{array}\!\right),
\end{equation}
where $H_{\mathrm{NLO}}$ is the NLO correction to the LO three-body force, which is again fit to the doublet $S$-wave $nd$ scattering length~\cite{Vanasse:2015}.  The dibaryons also receive a correction at NLO, which is given by the c.c.~space vector
\begin{equation}
\mathbf{R}_{1}(p,E)=
\left(\begin{array}{c}
\frac{(Z_{t}-1)}{2\gamma_{t}}\left(\gamma_{t}+\sqrt{\frac{3}{4}p^{2}-M_{N}E-i\epsilon}\right)\\[3mm]
\frac{(Z_{s}-1)}{2\gamma_{s}}\left(\gamma_{s}+\sqrt{\frac{3}{4}p^{2}-M_{N}E-i\epsilon}\right)
\end{array}\right).
\end{equation}

The $\nnlo$ kernel receives contributions from the two-body $SD$-mixing term, the $\nnlo$ correction to the LO three-body force, $H_{\nnlo}$, and a new energy dependent three-body force, $H_{2}^{\nnlo}$.  In c.c.~space these contributions give
\begin{align}
&\hspace{-.5cm}\left[\mathbf{K}^{J}_{2;L'S',LS}(q,p,E)\right]_{zx}=\frac{y \, y_{SD} \, M_{N}}{2}\left(Z_{SD}^{(1)}(J,L',S',L,S,x,z)\frac{1}{kp}\left[4p^{2}Q_{L}(a)+k^{2}Q_{L'}(a)\right]\right.\\\nonumber
&+Z_{SD}^{(1)}(J,L,S,L',S',z,x)\frac{1}{kp}\left[p^{2}Q_{L}(a)+4k^{2}Q_{L'}(a)\right]\\\nonumber
&\left.+\sum_{L''}\left[Z_{SD}^{(2)}(J,L',S',L,S,x,z,L'')+Z_{SD}^{(2)}(J,L,S,L',S',z,x,L'')\right]Q_{L''}(a)\right)\\\nonumber
&-\pi \left(H_{\nnlo}+\frac{4}{3}(M_{N}E+\gamma_{t}^{2})H_{2}^{(\nnlo)}\right)\delta_{L0}\delta_{LL'}\delta_{SS'}\delta_{S\nicefrac{1}{2}}(-1)^{x+z},
\end{align}
where the subscripts ``$x$" and ``$z$" refer to the matrix element in c.c.~space and $y=y_{s}=y_{t}$.  The value $x=1$ ($z=1$) corresponds to an initial (final) spin-triplet dibaryon state and $x=0$ ($z=0$) to an initial (final) spin-singlet dibaryon state.  Functions $Z_{SD}^{(1)}(\cdots)$ and $Z_{SD}^{(2)}(\cdots)$ are defined with 3$nj$-symbols, yielding
\begin{align}
&Z_{SD}^{(1)}(J,L',S',L,S,x,z)=2\,\sqrt{\widehat{x}\widehat{z}\widehat{(1-z)}\widehat{S}\widehat{S'}\widehat{L}}\,\sqrt{\frac{10}{3}}\,(-1)^{\frac{1}{2}+x+z+L+S+S'-J}\SJ{z}{\frac{1}{2}}{\frac{1}{2}}{1}{S'}{\frac{1}{2}}\\\nonumber
&\hspace{5cm}\times\SJ{2}{1}{x}{\frac{1}{2}}{S}{S'}\SJ{S'}{2}{S}{L}{J}{L'}\CG{L}{0}{2}{0}{L'}{0},
\end{align}
and
\begin{align}
&Z_{SD}^{(2)}(J,L',S',L,S,x,z,L'')=\sum_{L''}8\,\sqrt{\widehat{x}\widehat{z}\widehat{(1-z)}\widehat{S}\widehat{S'}\widehat{L}\widehat{L''}}\,(-1)^{z+L''+L}\SJ{1}{\frac{1}{2}}{\frac{1}{2}}{z}{S'}{\frac{1}{2}}\\[2mm]\nonumber
&\times\left(\NJ{\frac{1}{2}}{x}{S}{1}{L''}{L}{S'}{L'}{J}+\SJ{\frac{1}{2}}{1}{S'}{L'}{J}{L''}\SJ{L}{x}{L''}{\frac{1}{2}}{J}{S}+\frac{1}{3}(-1)^{1+L''+L}\frac{1}{\widehat{S}\widehat{L}}\delta_{LL'}\delta_{SS'}\right)\CG{L}{0}{1}{0}{L''}{0}\CG{L''}{0}{1}{0}{L'}{0},
\end{align}
where the hat is defined as $\widehat{x}=2x+1$.  At $\nnlo$ there is also a correction to the dibaryon propagators from $c_{0t}^{(1)}$, $c_{0s}^{(1)}$, and $\Delta_{s}^{(\nnlo)}$. In c.c.~space this is
\begin{equation}
\mathbf{R}_{2}(p,E)=-
\left(\begin{array}{cc}
\frac{(Z_{t}-1)^{2}}{2\gamma_{t}}\left(\gamma_{t}+\sqrt{\frac{3}{4}p^{2}-M_{N}E-i\epsilon}\right) &\\[3mm]
\frac{(Z_{s}-1)^{2}}{2\gamma_{s}}\left(\gamma_{s}+\sqrt{\frac{3}{4}p^{2}-M_{N}E-i\epsilon}\right) &-\,\, \frac{2}{3}\Delta_{s}^{(\nnlo)}D_{s} \! \left(E-\frac{p^{2}}{2M_{N}},p\right)
\end{array}\right) .
\end{equation}
The factor of 2/3 in front of $\Delta_{s}^{(\nnlo)}$ comes from the isospin projection.  $\Delta_{s}^{(\nnlo)}$ is only associated with the $nn$ spin-singlet dibaryon propagator, which contributes 2/3 to the total isospin invariant nucleon spin-singlet dibaryon amplitude ($t^{J;Nt\to Ns}_{n;L'S',LS}(k,p,E)$)	.

The $\ntlo$ kernel contains a correction to the LO three-body force, $H_{\ntlo}$, and a correction to the $\nnlo$ energy dependent three-body force, $H_{2}^{\ntlo}$, which gives
\begin{equation}
\hspace{-.5cm}\mathbf{K}^{J}_{3;L'S',LS}(q,p,E)=-\pi\left(H_{\ntlo}+\frac{4}{3}(M_{N}E+\gamma_{t}^{2})H_{2}^{\ntlo}\right)\delta_{L0}\delta_{LL'}\delta_{SS'}\delta_{S\frac{1}{2}}
\left(\!\!\begin{array}{rr}
1 & -1 \\
-1 & 1 
\end{array}\!\right).
\end{equation}
Griesshammer~\cite{Griesshammer:2005ga} argues that at $\ntlo$ there is a new divergence that mixes the Wigner-symmetric and Wigner-antisymmetric channels of the doublet $S$-wave,  but that the need for a new counter-term is suppressed due to the Pauli principle and should occur two orders higher.  However, Birse~\cite{Birse:2005pm} suggests that the Pauli principle is automatically included in the asymptotic analysis and therefore the need for a new counter-term will be at $\ntlo$.  We find that fitting $H_{\ntlo}$ to the doublet $S$-wave $nd$ scattering length and $H_{2}^{(\ntlo)}$ to the triton binding energy yields a $\ntlo$ correction to the doublet $S$-wave $nd$ scattering amplitude that is not properly renormalized.  This supports the claim made by Birse that a new three-body force that mixes Wigner-antisymmetric and Wigner-symmetric channels in the doublet $S$-wave will be necessary at $\ntlo$.  This will be addressed in future work. 

 An important additional contribution at $\ntlo$ comes from the two-body $P$-wave contact interactions.  These will be dealt with using a slightly different method.  There are also corrections to the dibaryon propagators at this order coming from $c_{0t}^{(2)}$ and $c_{0s}^{(2)}$, and the shape parameter corrections $c_{1t}$ and $c_{1s}$, which yield 
\begin{equation}
\mathbf{R}_{3}(p,E)=
\left(\begin{array}{c}
\left(\gamma_{t}+\sqrt{\frac{3}{4}p^{2}-M_{N}E-i\epsilon}\right)\left[\frac{(Z_{t}-1)^{3}}{2\gamma_{t}}+\rho_{t1}\left(\frac{3}{4}p^{2}-M_{N}E-\gamma_{t}^{2}\right)\right]\\[3mm]
\left(\gamma_{s}+\sqrt{\frac{3}{4}p^{2}-M_{N}E-i\epsilon}\right)\left[\frac{(Z_{s}-1)^{3}}{2\gamma_{s}}+\rho_{s1}\left(\frac{3}{4}p^{2}-M_{N}E-\gamma_{s}^{2}\right)\right]
\end{array}\right).
\end{equation}

\begin{figure}[hbt]
\begin{center}
\includegraphics[width=100mm]{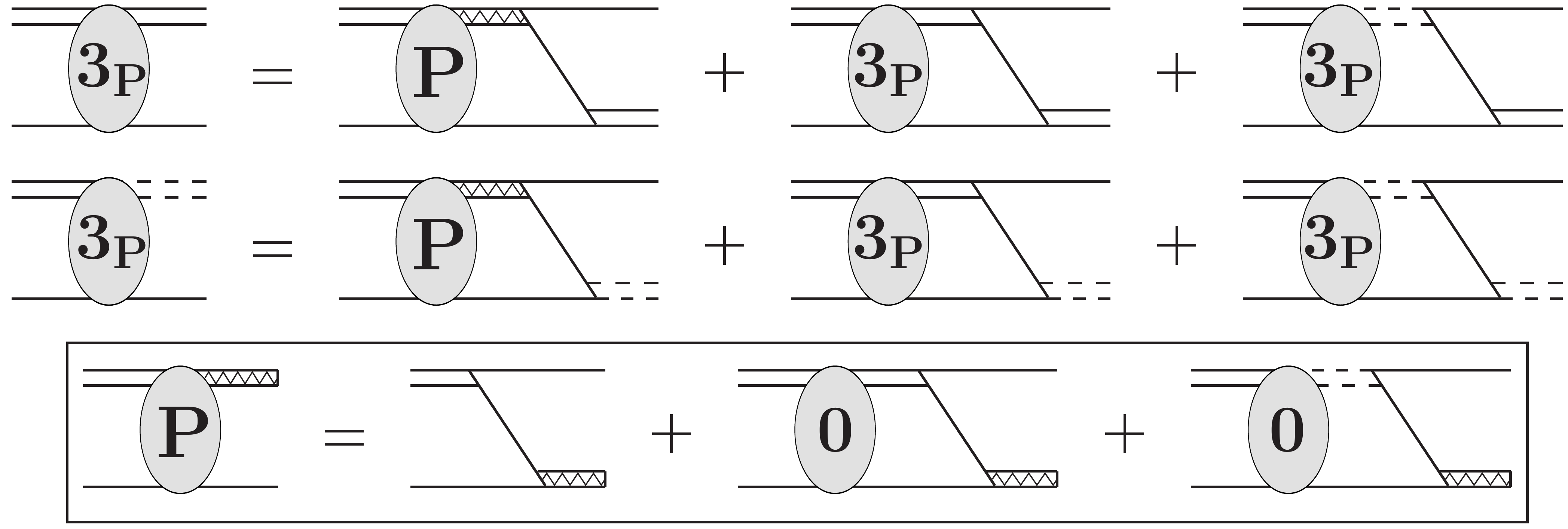}
\caption{\label{fig:TwoBodyPwave}Unboxed diagrams are the integral equations for the $\ntlo$ contribution to the $nd$ scattering amplitude from the two-body $P$-wave contact interactions.  The double lines with a zig-zag in the middle are the $P$-wave dibaryon propagator, given by $i/\Delta^{\wave{2R+1}{P}{J}}$.  The boxed diagrams represent the equation for the ``P'' amplitude used in the unboxed integral equations above.  The notation ``3$_{\rm P} $" in the oval indicates that this is a $\ntlo$ correction but one that only involves the two-body $P$-wave contributions.}
\end{center}
\end{figure}

The numerical solution of the integral equations is carried out by means of the Hetherington-Schick method \cite{Hetherington:1965zza,Aaron:1966zz,Ziegelmann}, which solves the equations along a contour in the complex plane.  Using the solution along the contour in the integral equations, the scattering amplitude along the real axis can be solved.  By rotating into the complex plane the fixed singularity of the deuteron pole is avoided as well as the branch cut singularities that occur above the deuteron breakup energy.  The methods in Refs.~\cite{Vanasse:2013sda,Vanasse:2015} allow the Hetherington-Schick method to be used to calculate diagrams with the full off-shell LO scattering amplitude without calculating the full off-shell scattering amplitude.  In order to calculate to large cutoffs and obtain sufficient numerical accuracy most mesh points are clustered for momenta with magnitude less than $\Lambda_{\not{\pi}}$.  For momenta with magnitude greater than $\Lambda_{\not{\pi}}$ far fewer mesh points are used since the amplitude is decaying as a power law \cite{Griesshammer:2005ga}.  In this way, calculations to large cutoffs can  be obtained using a reasonable number of mesh points, with a numerical accuracy of less than 1\%.  It is important to show convergence to large cutoffs because observables in this EFT should be independent of the cutoff.  Any deviations to the result as the cutoff becomes large indicates missing physics (such as a neglected counterterm).
\vskip .5cm
\leftline{\bf P-wave auxiliary field}
The contribution to the $\ntlo$ kernel from the two-body $P$-wave contact interaction is shown in the last box in Fig.~\ref{fig:N3LOndScatteringDiagrams}.  This diagram is  one-loop and can be solved analytically and projected out in an angular momentum basis.  However, the resulting forms for all partial waves are cumbersome in numerical calculations.  In order to circumvent this it is convenient to introduce a $P$-wave auxiliary field via the Lagrangian
%
%
\begin{align}
\label{eq:PwaveLag}
&\mathcal{L}_{2}^{P}=-{\Pc_{0A}^{\threePzero}}{}^{\dagger}\Delta^{(\threePzero)}\Pc_{0A}^{\threePzero}-{\Pc_{iA}^{\threePone}}{}^{\dagger}\Delta^{(\threePone)}\Pc_{iA}^{\threePone}-{\Pc_{iA}^{\threePtwo}}{}^{\dagger}\Delta^{(\threePtwo)}\Pc_{iA}^{\threePtwo}
-{\Pc_{i}^{\oneP}}{}^{\dagger}\Delta^{(\oneP)}\Pc_{i}^{\oneP}\\\nonumber
&\hspace{2cm}+\frac{1}{2}\sum_{J=0}^{2}y^{\wave{3}{P}{J}}\left[\CG{1}{i}{1}{j}{J}{k}\left(\Pc_{kA}^{\wave{3}{P}{J}}\right)^{\dagger}\hat{N}^{T}i\mathcal{O}_{jiA}^{(1,P)}\hat{N}+\mathrm{H.c.}\right]\\\nonumber
&\hspace{2cm}+\frac{1}{2}y^{\oneP}\left[\left(\Pc_{i}^{\oneP}\right)^{\dagger}\hat{N}^{T}i\mathcal{O}_{i}^{(0,P)}\hat{N}+\mathrm{H.c.}\right].
\end{align}
This approach is equivalent to using the two-body $P$-wave contact interactions in Eqs.~(\ref{eq:PwaveLagRupak}) and (\ref{eq:Pwave1P1}).  The $\ntlo$ contribution from the two-body $P$-wave contact terms is given by the coupled integral equations in Fig.~\ref{fig:TwoBodyPwave}. The ``P" amplitude is defined in the boxed region of Fig.~\ref{fig:TwoBodyPwave}, where the double line with a zig-zag represents a $P$-wave dibaryon propagator.  The $P$-wave dibaryon propagator is simply given by a constant, since the scattering volumes in the two-body $P$-waves are of natural size and are therefore perturbative and do not require resumming.  The constant term can be factored out of all numerical expressions and then reintroduced at the end to obtain the final expression.

The ``P" amplitude is given by 
\begin{equation}
\label{eq:Pamp}
t^{J({}^{2R+1}\!P_{z})}_{L'S',LS}(k,p,E)=\left[\mathbf{K}^{J({}^{2R+1}\!P_{z})}_{L'S',LS}(k,p,E)\right]_{1}+\mathbf{K}^{J({}^{2R+1}\!P_{z})}_{L'S',LS}(q,p,E)\otimes \mathbf{t}_{0;LS,LS}^{J}(k,q,E),
\end{equation}
where $R$=0 or 1, the quantity $z=0,1,2$ ($z=1$) for the ${}^{3}\!P_{J}$ (${}^{1}\!P_{1}$) two-body $P$-wave contact interaction channels (channel), and the kernel function $\mathbf{K}^{J({}^{2R+1}\!P_{z})}_{L'S',LS}(k,p,E)$ is a vector in c.c.~space.  For the first term on the right hand side only one element of the kernel in c.c.~space is taken since only states with an initial spin-triplet dibaryon are needed.  The kernel function for Eq.~(\ref{eq:Pamp}) is the tree level diagram in the box in Fig.~\ref{fig:TwoBodyPwave}. Projected onto a partial wave basis this yields
\begin{equation}
\left[\mathbf{K}_{L'S',LS}^{J({}^{3}\!P_{z})}(k,p,E)\right]_{x}=-\frac{M_{N} \, y \, y^{\wave{3}{P}{z}}}{4kp}\mathcal{Z}^{\wave{3}{P}{z}}\left(J,L',S',L,S,x,z\right)\left(2k \, Q_{L'}(a)+p \, Q_{L}(a)\right)
\end{equation}
for the ${}^{3}\!P_{J}$ coefficients, and
\begin{equation}
\left[\mathbf{K}_{L'S',LS}^{J({}^{1}\!P_{1})}(k,p,E)\right]_{x}=-\frac{M_{N} \, y \, y^{\oneP}}{4kp}\mathcal{Z}^{\oneP}\left(J,L',S',L,S,x\right)\left(2k \, Q_{L'}(a)+p \, Q_{L}(a)\right)
\end{equation}
for the ${}^{1}\!P_{1}$ coefficient.  Functions $\mathcal{Z}^{\wave{3}{P}{z}}(\cdots)$ and $\mathcal{Z}^{\wave{1}{P}{1}}(\cdots)$ are defined by
\begin{align}
&\mathcal{Z}^{\wave{3}{P}{z}}\left(J,L',S',L,S,x,z\right)=12(-1)^{\frac{3}{2}+S+S'+L-J}\sqrt{\widehat{x}\widehat{(1-x)}\widehat{z}\widehat{S}\widehat{S'}\widehat{L}}\,\SixJ{x}{\frac{1}{2}}{\frac{1}{2}}{1}{S}{\frac{1}{2}}\\\nonumber
&\hspace{1cm}\times\SixJ{\frac{1}{2}}{1}{S}{1}{S'}{z}\SixJ{S}{1}{S'}{L'}{J}{L}\SixJ{1-x}{\frac{1}{2}}{\frac{1}{2}}{1}{\frac{1}{2}}{\frac{1}{2}}\CG{L}{0}{1}{0}{L'}{0}
\end{align}
and
\begin{align}
\mathcal{Z}^{\oneP}\left(J,L',S',L,S,x\right)=(-1)^{\frac{3}{2}+L-J}\sqrt{\widehat{x}\widehat{(1-x)}\widehat{S'}\widehat{L}}\,\SixJ{S}{1}{S'}{L'}{J}{L}\delta_{S,\frac{1}{2}}\CG{L}{0}{1}{0}{L'}{0}.
\end{align}
As before, $x=1$ corresponds to an initial state spin-triplet dibaryon propagator and $x=0$ corresponds to an initial state spin-singlet dibaryon propagator.  The advantage of these kernel functions in the $P$-wave auxiliary field method is that they only contain Legendre functions of the second kind, which are already calculated to solve the LO $nd$ scattering amplitude.  Therefore, no additional work has to be done to calculate the values of these functions along the mesh points of our integral equations.  In the more conventional approach new functions involving $\arctan(\cdots)$ and $\log(\cdots)$  have to be calculated for each partial wave, and the higher the partial wave the more complicated the functional forms become.  The $P$-wave auxiliary field method is a far more transparent and numerically efficient means by which to calculate corrections from two-body $P$-wave contact terms to $nd$ scattering.  Using the ``P" amplitude the $\ntlo$ correction to the $nd$ scattering amplitude from the two-body $P$-wave contact interactions in Fig.~\ref{fig:TwoBodyPwave} is given by the coupled integral equations 
\begin{align}
&\mathbf{t}_{3_{\mathrm{P}};L'S',LS}^{J}(k,p,E)=\\\nonumber
&\hspace{2cm}\sum_{R=0}^{1} \ \sum_{z=|R-1|}^{R+1} \ \sum_{L'',S''}\frac{(-1)^{z}}{\Delta^{({}^{2R+1}\!P_{z})}}\left[\mathbf{K}^{J({}^{2R+1}\!P_{z})}_{L''S'',L'S'}(p,q,E)\right]^{T}\otimes t_{L''S'',LS}^{J({}^{2R+1}\!P_{z})}(k,q,E)\\\nonumber
&\hspace{5cm}+\mathbf{K}^{J}_{0;L'S',L'S'}(q,p,E)\otimes \mathbf{t}_{3_{\mathrm{P}};L'S',LS}^{J}(k,q,E).
\end{align}
The term $1/\Delta^{({}^{2R+1}\!P_{z})}$ is the $P$-wave dibaryon propagator, and can be removed from the integral equations by an appropriate renormalization.  In the inhomogeneous term we have used time reversal symmetry to write the kernel for a $P$-wave dibaryon going to an $S$-wave dibaryon in terms of the kernel for an $S$-wave dibaryon going to a $P$-wave dibaryon.  The factor of $(-1)^{z}$ is due to time reversal symmetry and comes from the fact that under time reversal
\begin{equation}
\CG{1}{i}{1}{j}{z}{k}\stackrel{T}{\longrightarrow}\CG{1}{-i}{1}{-j}{z}{-k}=(-1)^{z}\CG{1}{i}{1}{j}{z}{k}.
\end{equation}

Finally, the coefficients $y^{{}^{2S+1}\!P_{J}}$ and $\Delta^{({}^{2S+1}\!P_{J})}$ must be fit to $np$ scattering data.  This can be done by performing Gaussian integration on the $P$-wave auxiliary fields in Eq.~(\ref{eq:PwaveLag}) and matching to the coefficients in Eqs.~(\ref{eq:PwaveLagRupak}) and (\ref{eq:Pwave1P1}), or by performing a matching calculation using the Lagrangians of Eqs.~(\ref{eq:PwaveLag}), (\ref{eq:PwaveLagRupak}), and (\ref{eq:Pwave1P1}).  In order to match these coefficients the identities
\begin{equation}
\sum_{k}\CG{1}{i}{1}{j}{0}{0}\CG{1}{m}{1}{l}{0}{0}\to\frac{1}{3}\delta_{ij}\delta_{ml},
\end{equation}
\begin{equation}
\sum_{k}\CG{1}{i}{1}{j}{1}{k}\CG{1}{m}{1}{l}{1}{-k}(-1)^k\to\frac{1}{2}(\delta_{il}\delta_{jm}-\delta_{im}\delta_{jl}),
\end{equation}
and
\begin{equation}
\sum_{k}\CG{1}{i}{1}{j}{2}{k}\CG{1}{m}{1}{l}{2}{-k}(-1)^k\to\frac{1}{2}(\delta_{il}\delta_{jm}+\delta_{im}\delta_{jl}-\frac{2}{3}\delta_{ij}\delta_{ml})
\end{equation}
are used, where the indices on the left are spherical ($i,j,l,m=-1,0,1$) and those on the right Cartesian ($i,j,l,m=1,2,3$).  Matching coefficients yields
\begin{equation}
C^{{}^{3}\!P_{0}}=\frac{1}{3}\frac{(y^{{}^{3}\!P_{0}})^{2}}{\Delta^{({}^{3}\!P_{0})}}, \ \
C^{{}^{3}\!P_{1}}=-\frac{1}{2}\frac{(y^{{}^{3}\!P_{1}})^{2}}{\Delta^{({}^{3}\!P_{1})}}, \ \
C^{{}^{3}\!P_{2}}=\frac{1}{4}\frac{(y^{{}^{3}\!P_{2}})^{2}}{\Delta^{({}^{3}\!P_{2})}}, \ \
C^{{}^{1}\!P_{1}}=\frac{(y^{{}^{1}\!P_{1}})^{2}}{\Delta^{({}^{1}\!P_{1})}}.
\end{equation}
Then using Eq.~(\ref{eq:CPJnpfit}) the ratio of coefficients $y^{{}^{2S+1}\!P_{J}}$ and $\Delta^{({}^{2S+1}\!P_{J})}$ can be fit to $np$ scattering data.  A factor of $(y^{{}^{2S+1}\!P_{J}})^{2}/\Delta^{({}^{2S+1}\!P_{J})}$ can be removed from the $\ntlo$ correction to the $nd$ scattering amplitude by appropriate renormalization of the amplitude.  The calculated amplitude is then renormalized by this factor so that we can consider different values of $(y^{{}^{2S+1}\!P_{J}})^{2}/\Delta^{({}^{2S+1}\!P_{J})}$ without needing to recalculate the scattering amplitude.

Using the $P$-wave auxiliary field method is equivalent to starting with the one loop diagram for the two-body $P$-wave contact interaction diagram in Fig.~\ref{fig:N3LOndScatteringDiagrams} and projecting it in partial waves before carrying out the loop integration.  The resulting angular loop integration is trivial, leading to an integral over the magnitude of the loop momentum of an integrand of products of Legendre functions of the second kind. We find that using the $P$-wave auxiliary field allows us to treat this $\ntlo$ part of the correction in direct parallel with earlier corrections; note that the boxed diagrams in Fig.~\ref{fig:TwoBodyPwave} mimic those found in the $S$-wave case.  The $P$-wave auxiliary field method effectively trades performing a one loop diagram analytically for a  tree level diagram that is then used to numerically reproduce the contributions from the one loop diagram.
\section{\label{sec:observables} Observables}

In the on-shell limit the scattering amplitudes become the transition matrix $M$ defined in the partial wave basis by
\begin{equation}
M_{L'S',LS}^{J}(k)=Z_{\mathrm{LO}}t_{L'S',LS}^{J;Nt\to Nt}(k,k,E).
\end{equation}
In order to calculate polarization observables the transition matrix in the spin-basis is needed.  This is related to the transition matrix in the partial wave basis via 
\begin{align}
&M_{m_{1}',m_{2}';m_{1},m_{2}}=\sqrt{4\pi}\sum_{J}\sum_{L,L'}\sum_{S,S'}\sum_{m_{S},m_{S}'}\sum_{m_{L}'}\sqrt{\widehat{L}}\,\CG{1}{m_{1}}{\frac{1}{2}}{m_{2}}{S}{m_{S}}\CG{1}{m_{1}'}{\frac{1}{2}}{m_{2}'}{S'}{m_{S}'}\\\nonumber
&\hspace{5cm}\times\CG{L}{0}{S}{m_{S}}{J}{M}\CG{L'}{m_{L}'}{S'}{m_{S}'}{J}{M}Y_{L'}^{m_{L}'}(\theta,\phi)M_{L'S',LS}^{J}(k),
\end{align}
where $m_{1}$ ($m_{1}'$) is the initial (final) deuteron spin component in the $z$-direction and $m_{2}$ ($m_{2}'$) the initial (final) neutron spin component in the $z$-direction.  The unpolarized cross section is given by summing over all final spins and averaging over all initial spins of the square of the transition matrix, yielding
\begin{equation}
\frac{d\bar{\sigma}}{d\Omega}(\theta)=\frac{1}{6}\left(\frac{M_{N}}{3\pi}\right)^{2}\sum_{m_{1},m_{2}}\sum_{m_{1}',m_{2}'}\left|M_{m_{1}',m_{2}';m_{1},m_{2}}\right|^{2},
\end{equation}
where the bar over $\sigma$ denotes that it is unpolarized.

Polarizing the initial neutron leads to three different polarization observables denoted $A_{y}$, $A_{x}$, and $A_{z}$. These correspond to polarizing the neutron along each of the respective axes.  In the Madison conventions~\cite{darden1971polarization} the $z$-direction is defined by the momentum of the incoming beam, $\hat{\mathbf{k}}_{i}$, and the $y$-direction by $\hat{\mathbf{k}}_{i}\times\hat{\mathbf{k}}_{f}$, where $\vect{k}_{i}$ ($\vect{k}_{f}$) is the incoming (outgoing) momentum of the neutron.  The polarization observables $A_{x}$ and $A_{z}$ are parity-violating observables and $A_{z}$ has been considered elsewhere in \EFT~\cite{Arani:2011if}.  The cross section due to a transversely polarized neutron beam is given by
\begin{equation}\label{sigAy}
\frac{d\sigma}{d\Omega}(\theta,\phi)=\frac{d\bar{\sigma}}{d\Omega}(\theta)\Big{[}1-A_{y}(\theta)\sin(\phi)\Big{]},
\end{equation}
where $\phi$ is the azimuthal angle, $\hat{\mathbf{k}}_{i}$ defines the $z$-direction, and the direction of polarization defines the $x$-axis.  An analyzing power $A_{y}(\theta,\phi)$ can be derived from the transition matrix in the spin-basis using density matrix techniques~\cite{Ohlsen:1972zz,glockle2012quantum}, which yield 
\begin{equation}
A_{y}(\theta,\phi)=\frac{\displaystyle\sum_{m_{1}}\sum_{m_{1}',m_{2}'}2\,\mathrm{Im}\left[M_{m_{1}',m_{2}';m_{1},\frac{1}{2}} \, M^{*}_{m_{1}',m_{2}';m_{1},-\frac{1}{2}}\right]}{\displaystyle\sum_{m_{1},m_{2}}\sum_{m_{1}',m_{2}'}\left|M_{m_{1}',m_{2}';m_{1},m_{2}}\right|^{2}}.
\end{equation}
The resulting form will contain a $\sin(\phi)$ that can be factored out to give the expression for $A_{y}(\theta)$ in Eq.~(\ref{sigAy}).

Polarizing the initial deuteron gives four polarization observables $iT_{11}$, $T_{20}$, $T_{21}$, and $T_{22}$.  Other polarization observables exist but are related by rotational symmetry or are parity-violating or violate time-reversal symmetry.  The differential scattering cross section in terms of these polarization observables is given by~\cite{darden1971polarization}
\begin{align}
&\frac{d\sigma}{d\Omega}(\theta,\phi)=\frac{d\bar{\sigma}}{d\Omega}(\theta)\Big[1+2\mathrm{Re}(it_{11})iT_{11}(\theta)\sin(\phi)+t_{20}T_{20}(\theta)\\\nonumber
&\hspace{3cm}+2\mathrm{Re}(t_{21})T_{21}(\theta)\cos(\phi)+2\mathrm{Re}(t_{22})T_{22}(\theta)\cos(2\phi)\Big],
\end{align}
where $t_{11}$, $t_{20}$, $t_{21}$, and $t_{22}$ are numbers giving the amount of respective polarization.  Using density matrix techniques the vector polarization $iT_{11}(\theta,\phi)$ is given by
\begin{equation}
iT_{11}(\theta,\phi)=-\sqrt{\frac{3}{2}} \ \frac{\displaystyle\sum_{m_{2}}\sum_{m_{1}',m_{2}'}\,\mathrm{Im}\left[M_{m_{1}',m_{2}';-1,m_{2}}M^{*}_{m_{1}',m_{2}';0,m_{2}}+M_{m_{1}',m_{2}';0,m_{2}}M^{*}_{m_{1}',m_{2}';1,m_{2}}\right]}{\displaystyle\sum_{m_{1},m_{2}}\sum_{m_{1}',m_{2}'}\left|M_{m_{1}',m_{2}';m_{1},m_{2}}\right|^{2}},
\end{equation}
where $\sin(\phi)$ can be factored out to give $iT_{11}(\theta)$.  The tensor polarizations using density matrix techniques are given by
\begin{equation}
T_{20}(\theta)=\frac{1}{\sqrt{2}} \ \frac{\displaystyle\sum_{m_{2}}\sum_{m_{1}',m_{2}'}\left\{\left|M_{m_{1}',m_{2}';1,m_{2}}\right|^{2}-2\left|M_{m_{1}',m_{2}';0,m_{2}}\left|^{2}+\right|M_{m_{1}',m_{2}';-1,m_{2}}\right|^{2}\right\}}{\displaystyle\sum_{m_{1},m_{2}}\sum_{m_{1}',m_{2}'}\left|M_{m_{1}',m_{2}';m_{1},m_{2}}\right|^{2}},
\end{equation}
\begin{equation}
T_{21}(\theta,\phi)=-\sqrt{\frac{3}{2}} \ \frac{\displaystyle\sum_{m_{2}}\sum_{m_{1}',m_{2}'}\,\mathrm{Re}\left[M_{m_{1}',m_{2}';0,m_{2}}\left(M^{*}_{m_{1}',m_{2}';1,m_{2}}-M^{*}_{m_{1}',m_{2}';-1,m_{2}}\right)\right]}{\displaystyle\sum_{m_{1},m_{2}}\sum_{m_{1}',m_{2}'}\left|M_{m_{1}',m_{2}';m_{1},m_{2}}\right|^{2}},
\end{equation}
and
\begin{equation}
T_{22}(\theta,\phi)=\sqrt{3} \ \frac{\displaystyle\sum_{m_{2}}\sum_{m_{1}',m_{2}'}\mathrm{Re}\left[M_{m_{1}',m_{2}';1,m_{2}}M_{m_{1}',m_{2}';-1,m_{2}}^{*}\right]}{\displaystyle\sum_{m_{1},m_{2}}\sum_{m_{1}',m_{2}'}\left|M_{m_{1}',m_{2}';m_{1},m_{2}}\right|^{2}},
\end{equation}
where again the $\phi$ dependence can be factored out leaving only $\theta$ dependence.  The polarization observables can also be derived using the techniques in Ref.~\cite{Fukukawa:2011wu}, which has the advantage of writing the angular dependence in terms of Legendre polynomials, where their coefficients are given by the scattering amplitudes.  This gives analytical insight into how the shape of polarization observables are related to the scattering amplitudes.

\section{\label{sec:results} Results}

\begin{figure}[hbt]
	\begin{center}
	\begin{tabular}{cc}
		\hspace{-1cm}\includegraphics[width=90mm]{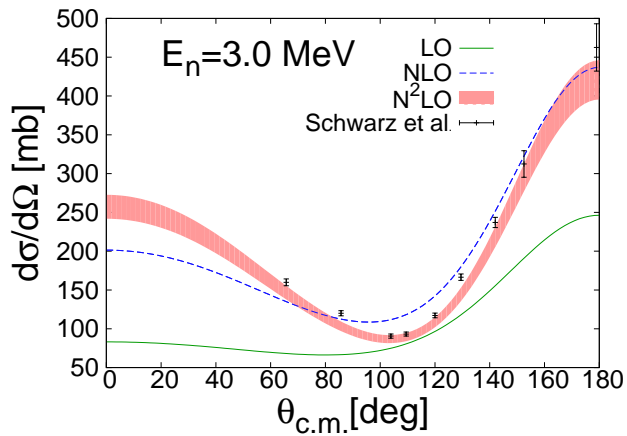} &
	\end{tabular}
	\end{center}

\caption{\label{fig:sigma}$nd$ scattering cross section for $E_{n}=3.0$~MeV  with experimental data from Schwarz et al.~\cite{Schwarz19831}.  The LO prediction (without theoretical errors) is the solid green line, the dashed blue line the NLO prediction (without theoretical errors), and the solid red band the $\nnlo$ prediction with a 6\% error estimate.}
\end{figure}

The \EFT differential scattering cross section up to $\nnlo$ at a neutron lab energy of $E_{n}=3.0$~MeV is shown in  Fig.~\ref{fig:sigma} and compared with data from Schwarz et al.~\cite{Schwarz19831}.  The solid green line is the LO prediction, the dashed blue line the NLO prediction, and the red band the $\nnlo$ prediction with a 6\% error estimate from the \EFT power counting.  Relatively good agreement at $\nnlo$ is observed with respect to the experimental data, and the minimum at $\nnlo$ coincides with that of the available experimental data.  Theoretical errors are not shown on the LO and NLO results, but we see that the \EFT treatment converges on the data.  The \EFT power counting predicts a naive error estimate of $(Q/\Lambda_{\not{\pi}})^{n}\!\!\sim\!\!(1/3)^{n}$ for the N$^{n}$LO scattering amplitudes, where $Q\!\sim\!\gamma_{t}$ and $\Lambda_{\not{\pi}}\!\sim\! m_{\pi}$.
The $\ntlo$ cross section is not shown. The doublet $S$-wave channel at $\ntlo$ is not renormalized by the three-body forces $H_{\ntlo}$ and $H_{2}^{\ntlo}$ alone.  Another three-body force is missing, one that likely mixes the Wigner-symmetric and Wigner-antisymmetric channels. This issue will be addressed in future work.

\begin{figure}[Hbt!]
	\begin{center}
	\begin{tabular}{cc}
	\hspace{-1cm}\includegraphics[width=90mm]{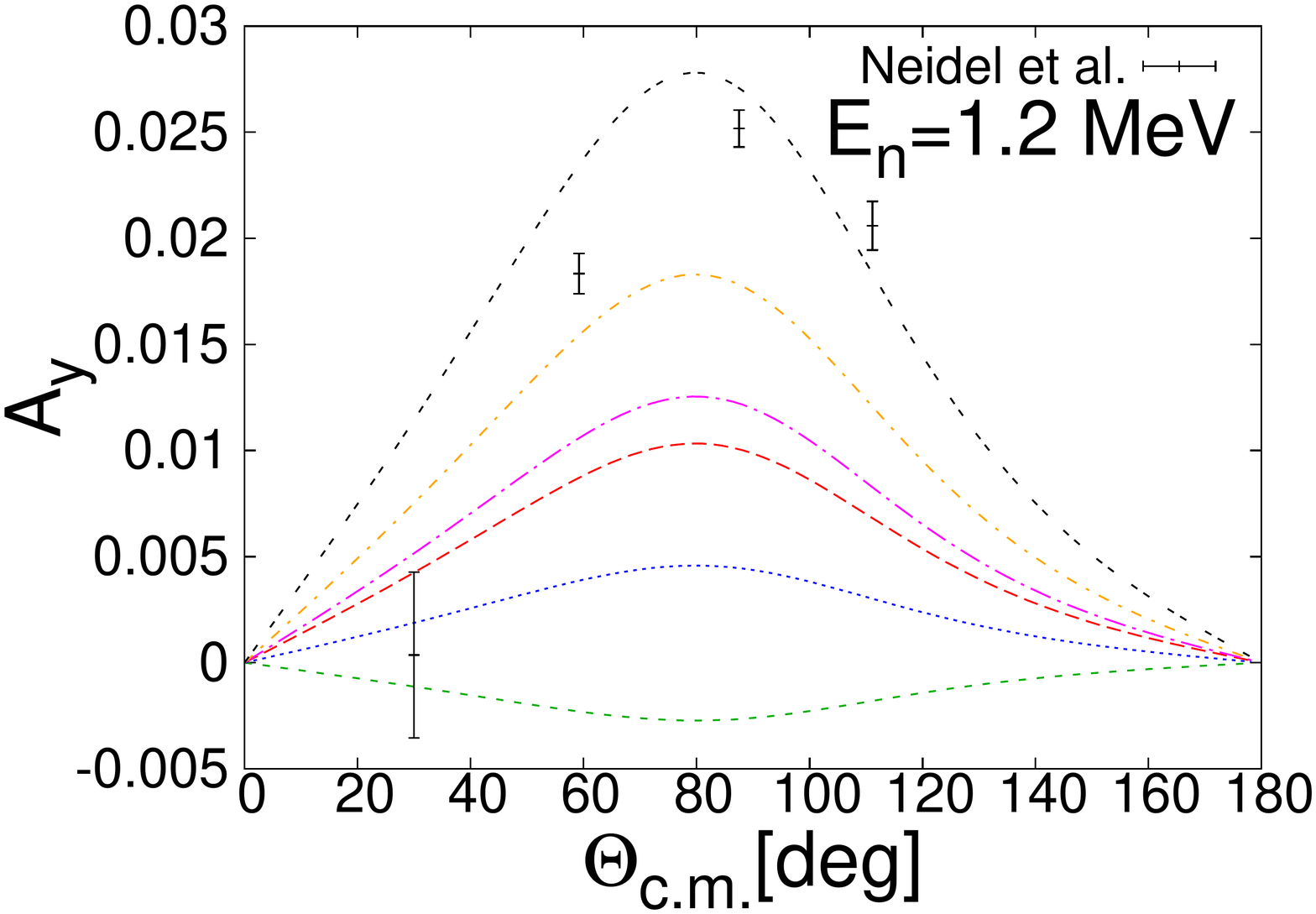} &
	\hspace{-1cm}\includegraphics[width=90mm]{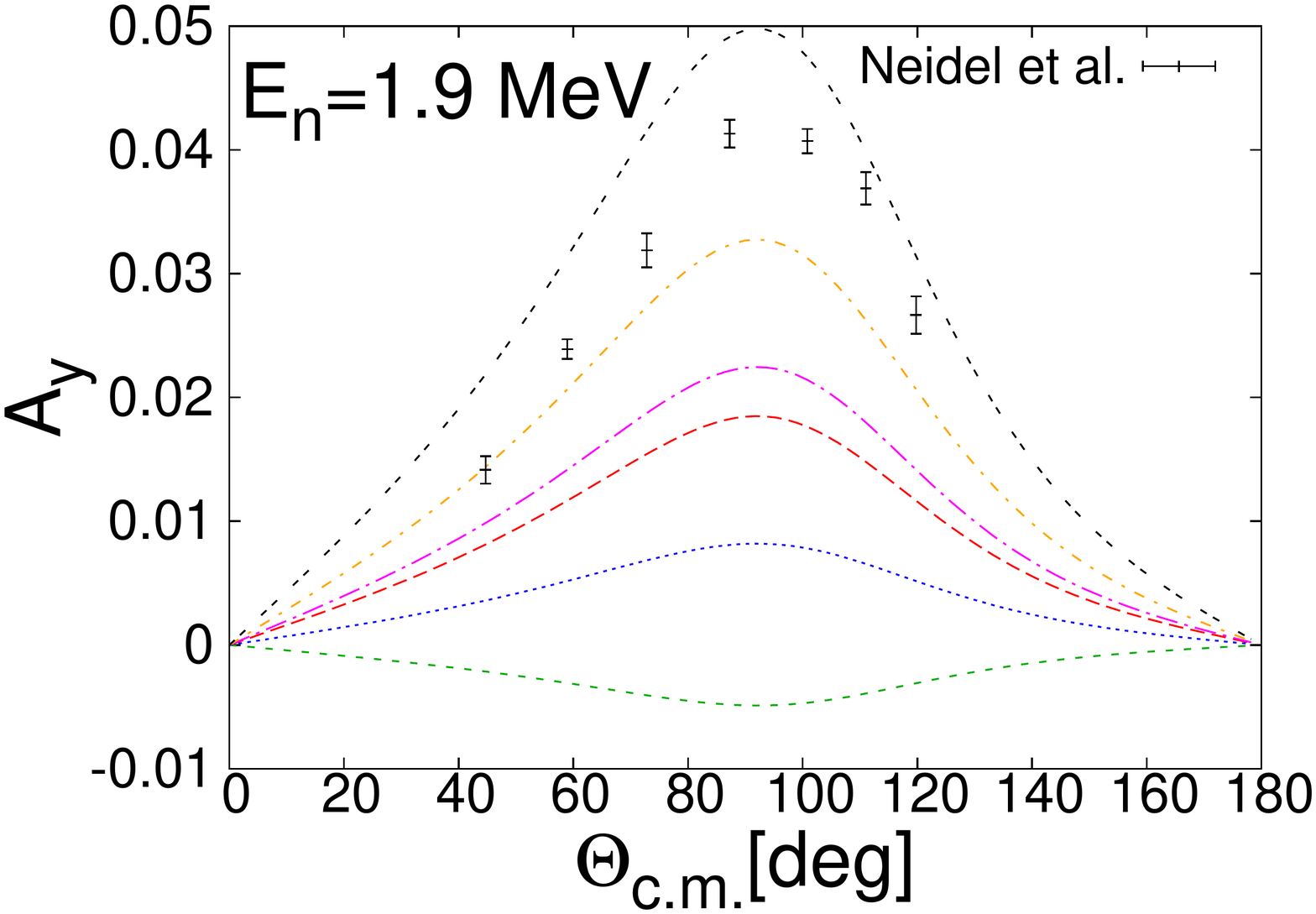} \\
	\end{tabular}

	\vspace{-.5cm}
	
	\includegraphics[width=90mm]{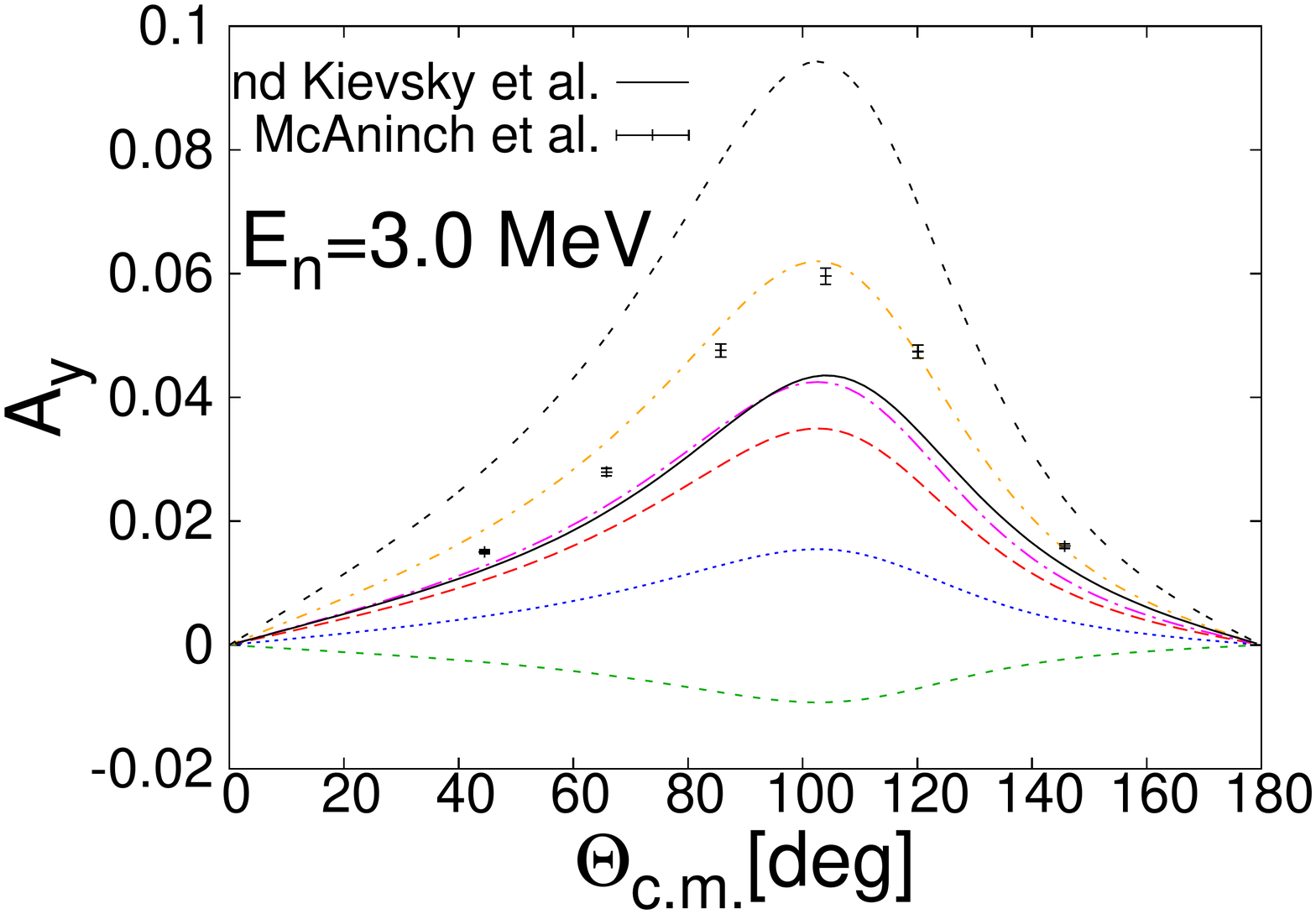}
	\end{center}

\caption{\label{fig:Aybands}The dashed lines are \EFT results for $A_{y}$ for several sets of $C^{{}^{3}\!P_{J}}$ coefficients varied by 15\% around their central values.   Top Left: $E_{n}=1.2$~MeV, experimental data from Neidel et al.~\cite{neidel2003new}. Top Right: $E_{n}=1.9$~MeV, experimental data from Neidel et al.~\cite{neidel2003new}. Bottom: $E_{n}=3.0$~MeV, the solid line is a PMC calculation from Kievsky et al. using AV-18+UR~\cite{Kievsky:1996ca}, with experimental data from McAninch et al.~\cite{McAninch:1994zz}. In the following, ``$+$" stands for 15 percent above central values given in Eq.~(\ref{eq:CPJnpfit}); ``$0$" is at central value; and ``$-$" is 15 percent below central value.  The coefficient values ($C^{{}^{3}\!P_{0}}$,$C^{{}^{3}\!P_{1}}$,$C^{{}^{3}\!P_{2}}$) used to produce the curves shown are (from lowest \EFT curve to highest \EFT curve on the plots): big dots (green)=$(+,-,+)$; small dots (blue)=$(+,0,+)$; long dash (red)=$(0,0,0)$; long-dash-dot (purple) = $(0,0,+)$; short-dash-dot (orange) = $(-,0,0)$; double-dot (black) = $(-,+,+)$.}
\end{figure}

The results of the \EFT calculation for the vector analyzing power, $A_{y}$, at neutron lab energies of $E_{n}=1.2$,  $1.9$, and 3.0~MeV are shown in Fig.~\ref{fig:Aybands}.  At $E_{n}=3.0$~MeV the solid black line is from a PMC using the AV-18 and Urbana (UR) potential with the hyperspherical harmonics technique~\cite{Kievsky:1996ca}. The experimental data for $A_{y}$ at $E_{n}=1.2$ and 1.9~MeV is from Neidel et al.~\cite{neidel2003new} and  those at $E_{n}=3.0$~MeV are from McAninch et al.~\cite{McAninch:1994zz}.  The peak of $A_{y}$ is primarily determined by the minimum of the cross section. All of the polarization observables are given as a ratio of cross sections in which the unpolarized cross section is in the denominator.  In a strictly perturbative $\ntlo$ calculation of polarization observables the denominator should be expanded and the unpolarized cross section need only be calculated to NLO, because the numerator of all polarization observables starts at $\nnlo$ from the first non-zero contribution from two-body $SD$-mixing.  However, expanding the denominator puts the minimum at the (experimentally) wrong place and  the maximum of $A_{y}$ qualitatively at the wrong place.  Therefore, for these results the perturbative cross section up to $\nnlo$ is kept in the denominator for polarization observables without the denominator being further expanded.\footnote{The different peak position of $A_{y}$ between treating the denominator strictly perturbatively or resumming higher order contributions can be considered an estimate of the \EFT error.  Therefore a more accurate assessment of the \EFT error of $A_{y}$ would not only include the error in the magnitude of $A_{y}$, but also the error in its peak position.  However, the magnitude of $A_{y}$ primarily depends on the two-body $P$-wave contact interactions and the peak position is independent of these contact interactions.  Thus choosing the peak position of $A_{y}$ to be near the experimental values, the error in the magnitude of $A_{y}$ is then separately addressed.}

While the position of the peak of $A_{y}$ depends primarily on the minimum of the cross section, its magnitude depends primarily on  the two-body $P$-wave contact interaction $C^{{}^{3}\!P_{J}}$ terms.  The contribution to $A_{y}$ from the $SD$-mixing term is negligible at about three orders of magnitude smaller than the contribution from the two-body $P$-wave contact interactions. In Fig.~\ref{fig:Aybands} the different dashed curves correspond to different choices for the $C^{{}^{3}\!P_{J}}$ coefficients.  These coefficients are fit to the Nijmegen $P$-wave phase shifts for $np$ scattering, yielding the central values in Eq.~(\ref{eq:CPJnpfit}).  But we expect a substantial \EFT theoretical error associated with this fit because these terms are the leading \EFT terms contributing to $NN$ $P$-wave scattering.  Therefore we vary each of the three $C^{{}^{3}\!P_{J}}$ coefficients by 15 percent from their central values.  Representative curves are shown in Fig.~\ref{fig:Aybands}.  The most obvious conclusion is that when appropriate theoretical errors are applied to the $C^{{}^{3}\!P_{J}}$ coefficients, a large range of $A_y$'s can be accommodated.  This motivates us to go to higher order.

\begin{figure}[Hbt!]
\begin{center}
	\begin{center}
	\begin{tabular}{cc}
	\hspace{-1cm}\includegraphics[width=90mm]{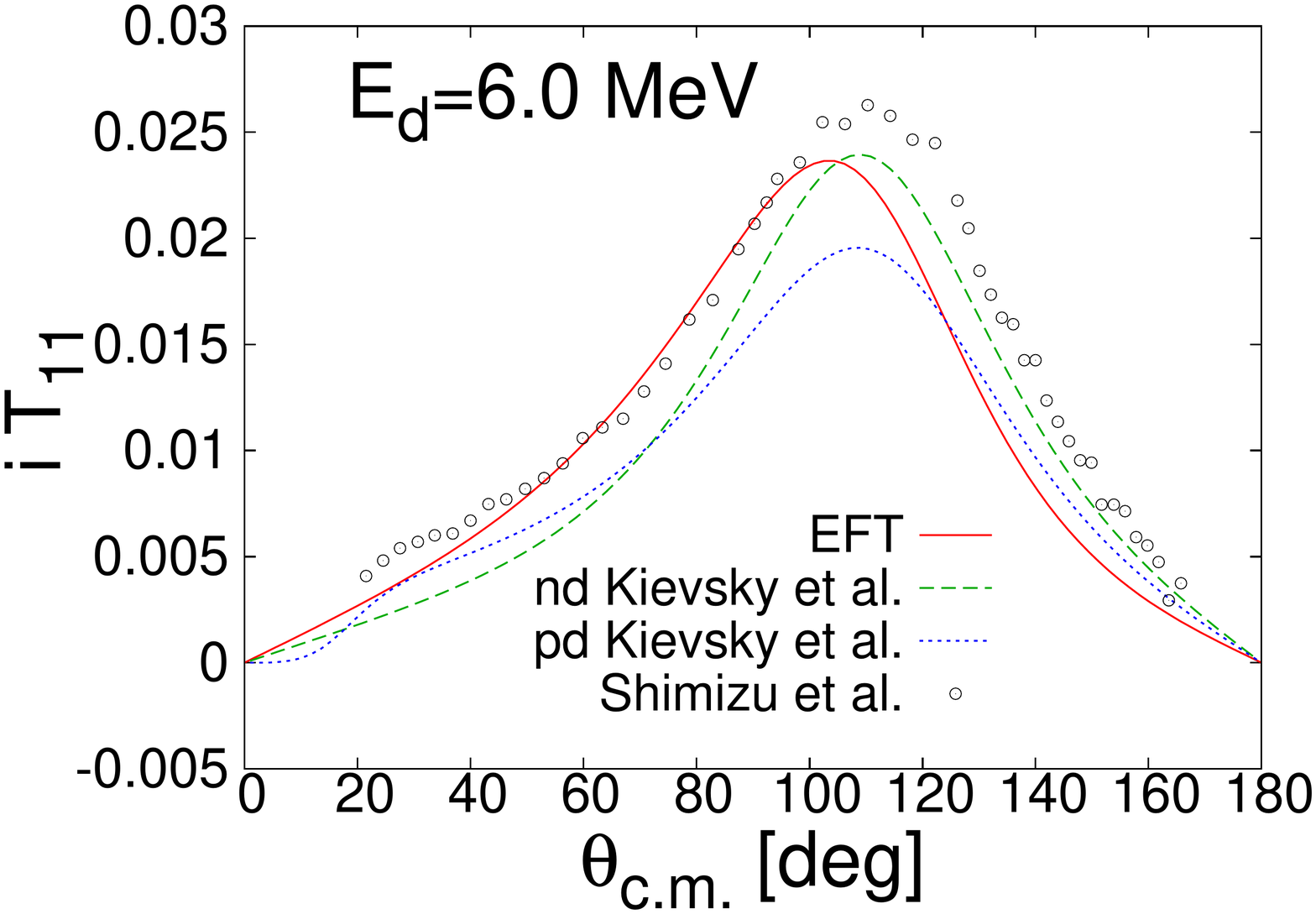} &
	\hspace{-1cm}\includegraphics[width=90mm]{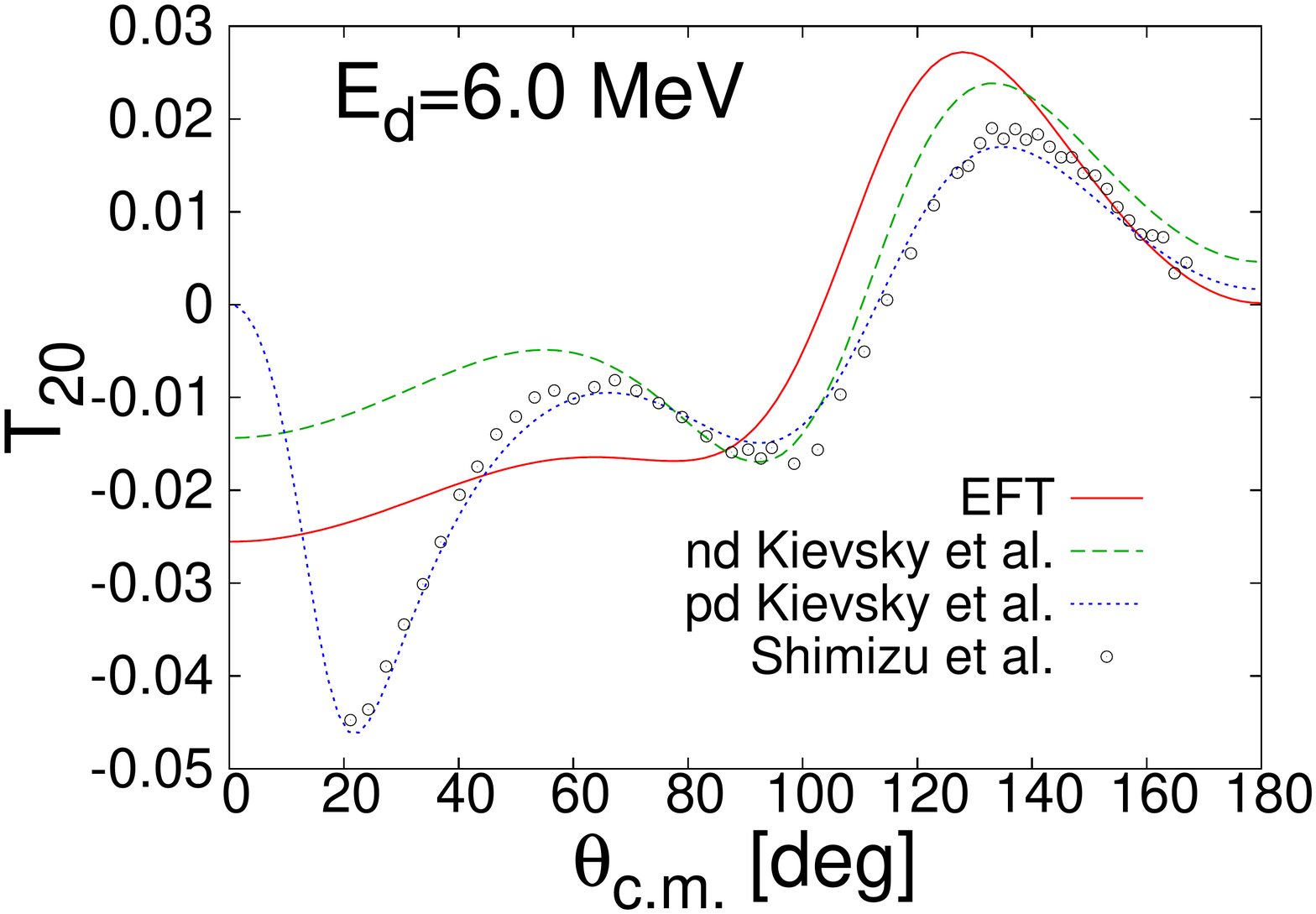} \\[-.5cm]

	\hspace{-1cm}\includegraphics[width=90mm]{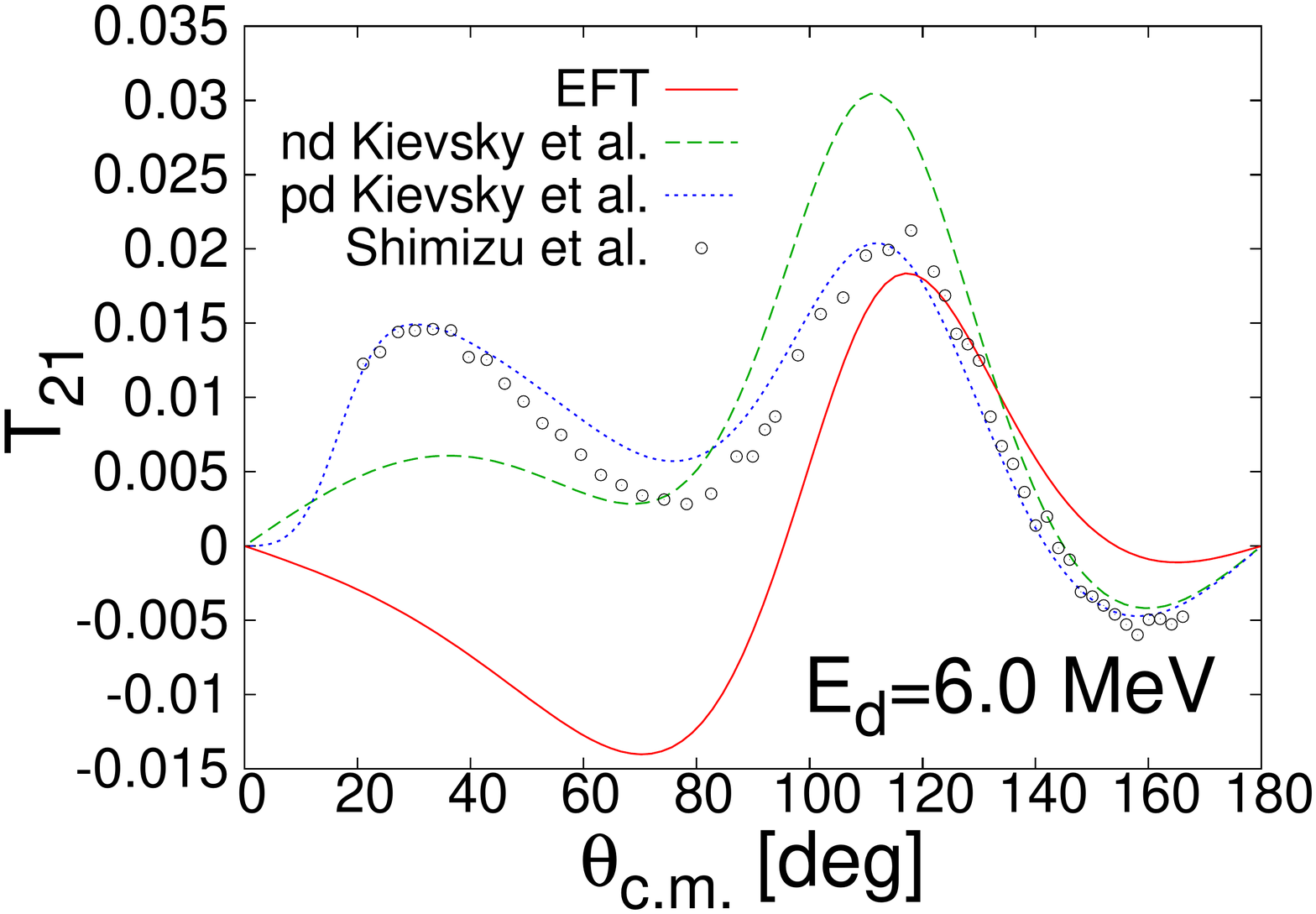} &
	\hspace{-1cm}\includegraphics[width=90mm]{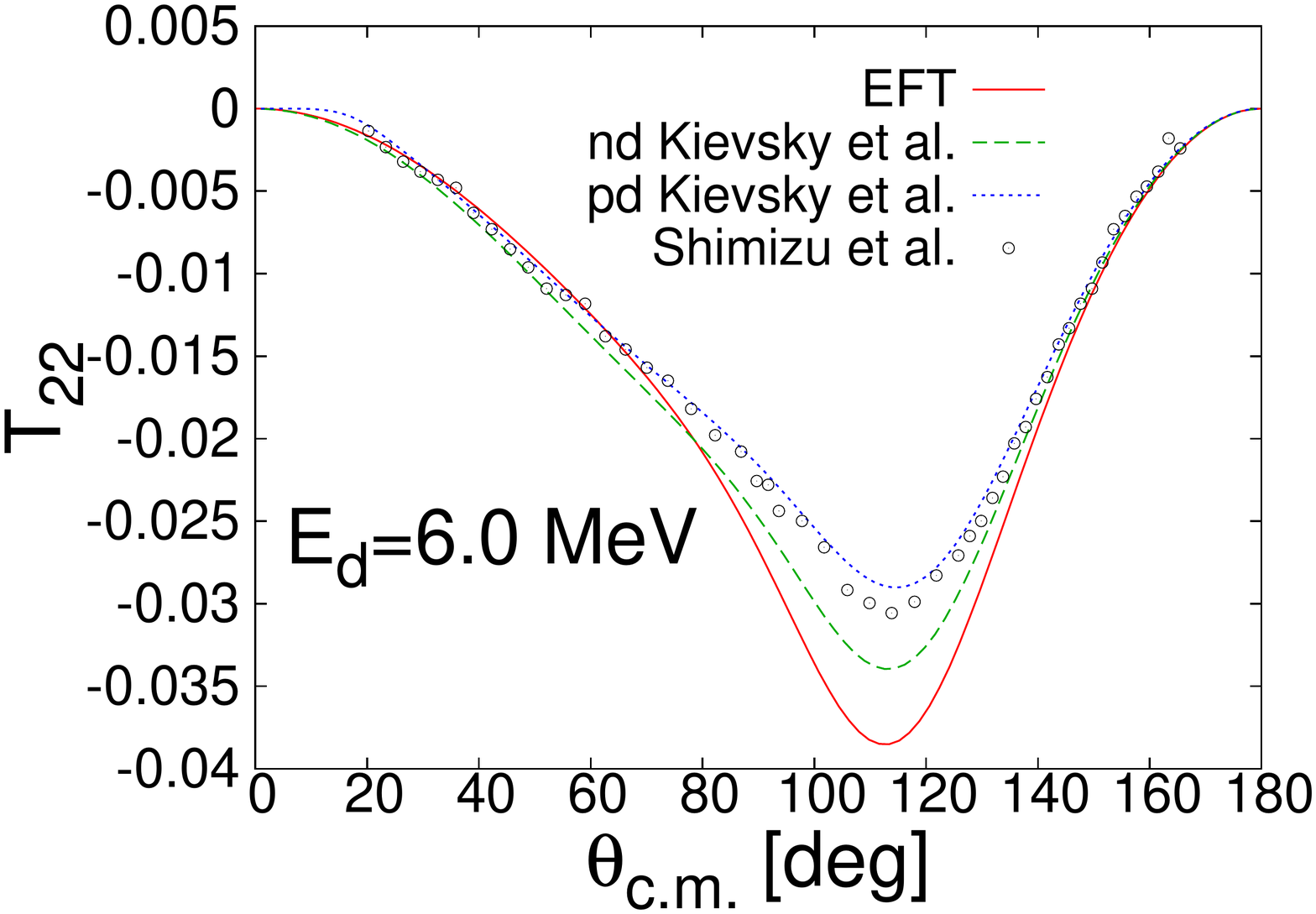}
	\end{tabular}
	\end{center}


\caption{\label{fig:Tnn}The solid red line is the \EFT prediction (without theoretical error bars) for deuteron polarization observables in $nd$ scattering, the dashed green line PMC calculations using AV18+UR for $nd$ scattering~\cite{Kievsky:1996ca}, and the dotted blue line PMC calculations using AV18+UR for $pd$ scattering~\cite{Kievsky:1996ca}. All experimental data is for $pd$ scattering from Shimizu et al.~\cite{Shimizu:1995zz} at a laboratory deuteron energy of $E_{d}=6.0$~MeV.}
\end{center}
\end{figure}

While the curves shown in Fig.~\ref{fig:Aybands} are just a representative sample, we note the following scaling:  As $C^{{}^{3}\!P_{0}}$ increases from the central value in Eq.~(\ref{eq:CPJnpfit}), with the other coefficients held constant, $A_y$ decreases substantially at all energies. $A_y$ is most sensitive to variations of this coefficient about its central value when the other two coefficients are held constant within their allowed variation.  Next in sensitivity are changes in $C^{{}^{3}\!P_{1}}$, and then changes in $C^{{}^{3}\!P_{2}}$ (again with other coefficients held constant within their allowed variation).  But for the $J=1,2$ values, $A_y$ increases as these coefficients increase. The contributions of the $C^{{}^{3}\!P_{J}}$ coefficients to $A_{y}$ are not independent since if the $C^{{}^{3}\!P_{J}}$ coefficients are all equal their total contribution to $A_{y}$ is zero.

All of the $A_{y}$ observables are calculated at a cutoff of $\Lambda=10^{6}$~MeV.  It is necessary to use a large cutoff, because the  three-body ${}^{4}\!P_{J}$, ${}^{2}\!P_{J}$, and ${}^{4}\!P_{J}\to{}^{2}\!P_{J}$ channels go asymptotically like $q^{-0.545\cdots}$~\cite{Griesshammer:2005ga} and as a function of $\Lambda$ converge slowly.\footnote{For the ${}^{2}\!P_{J}$ channel the leading asymptotic behavior comes from the Wigner-antisymmetric piece, which is equivalent to the asymptotic behavior of the ${}^{4}\!P_{J}$ channel \cite{Griesshammer:2005ga}.}  A  new three-body counter-term for the ${}^{4}\!P_{J}$, ${}^{2}\!P_{J}$, and ${}^{4}\!P_{J}\to{}^{2}\!P_{J}$ channels will be needed at order $\mathrm{N}^{3.5}\mathrm{LO}$~\cite{Griesshammer:2005ga} and is necessary for a $\mathrm{N}^{4}\mathrm{LO}$ calculation.  In addition, the channel ${}^{2}\!S_{\nicefrac{1}{2}}\to{}^{4}\!D_{\nicefrac{1}{2}}$ goes asymptotically like $q^{-0.105...+i1.00624\cdots}$.  Despite the slow rate of convergence, varying this channel within its cutoff variation was found to have almost no effect on the observables studied here at the order to which we are working.  However, given that a three-body force occurs in this channel at order $\mathrm{N}^{3.1}\mathrm{LO}$~\cite{Griesshammer:2005ga}, a new three body force must be included in a $\mathrm{N}^{4}\mathrm{LO}$ calculation.  All other channels converge much faster and are well converged at cutoffs of a few thousand MeV. 

The deuteron polarization observables are given in Fig.~\ref{fig:Tnn}, where the solid red line is the \EFT prediction (not including theoretical errors) using the central values for $C^{{}^{3}\!P_{J}}$ coefficients given in Eq.~(\ref{eq:CPJnpfit}), the long-dashed green line a PMC for $nd$ using AV-18+UR~\cite{Kievsky:1996ca}, and the dotted blue line a PMC for $pd$ using AV-18+UR~\cite{Kievsky:1996ca}.  All of the data shown is for $pd$ scattering from Shimizu et al.~\cite{Shimizu:1995zz} at a laboratory deuteron energy of $E_{d}=6.0$~MeV.  The results for $nd$ scattering  should roughly agree with those of $pd$ scattering at higher energies and backward angles where Coulomb effects are less important.  Rough qualitative agreement is observed  for all polarization observables.  The contribution from $SD$-mixing is significant for all polarization observables except $iT_{11}$, where the $SD$-mixing contribution is three orders of magnitude smaller than the two-body $P$-wave contact interaction contributions.  In fact, $iT_{11}$ changes in the same manner that $A_y$ changes when the  two-body $P$-wave coefficients are varied.  The contribution from the two-body $P$-wave contact interactions for $T_{20}$ is roughly the same size as $SD$-mixing contributions, negligible for $T_{21}$, and about two orders of magnitude smaller than $SD$-mixing  for $T_{22}$.  

\section{\label{sec:conclusion} Conclusion}

Using the techniques in Refs.~\cite{Vanasse:2013sda,Vanasse:2015}, polarization observables in $nd$ scattering have been calculated to $\ntlo$ in \EFT.  The polarization observables in \EFT receive non-zero contributions from the two-body $SD$-mixing term and the two-body $P$-wave contact interactions, $C^{{}^{3}\!P_{J}}$. Contributions from the two-body $P$-wave contact interactions were calculated by the introduction of a $P$-wave auxiliary field.  This approach leads to great analytical and numerical simplifications in the calculation of the two-body $P$-wave contact interaction contributions to $nd$ scattering amplitudes at $\ntlo$, and will be useful for higher order \EFT calculations and halo EFT calculations as well.  We find that the $A_{y}$ and $iT_{11}$ polarization observables are dominated by contributions from the two-body $P$-wave contact interactions while $SD$-mixing contributions are three orders of magnitude smaller.  The $T_{20}$ observable receives roughly equal contributions from $SD$-mixing and two-body $P$-wave contact interactions, $T_{21}$ only has contributions from $SD$-mixing, and $T_{22}$ primarily receives contributions from $SD$-mixing with contributions from the two-body $P$-wave contact interactions being two orders of magnitude smaller.

Deuteron polarization observables for $nd$ scattering were compared with available $pd$ data and PMC.  Rough qualitative agreement is found at larger angles where effects from the Coulomb interaction are less important.  The nucleon vector analyzing power $A_{y}$ was calculated at $E_{n}=1.2$, 1.9, and $3.0$~MeV and was found to vary widely given the theoretical error associated with the two-body $P$-wave coefficients. However, the position of the $A_{y}$ peak is predicted well in \EFT.  Significantly, we find that we can account for all the low energy $A_y$ data considered here so long as we allow the extracted coefficients $C^{{}^{3}\!P_{J}}$ to vary within their expected \EFT theoretical errors.  Hence there is no disagreement at the moment between \EFT and experiment within theoretical error.  But this motivates us to pursue a higher order calculation.

In calculating the $nd$ scattering amplitudes the cutoff was taken to be $\Lambda=10^{6}$~MeV.  A large cutoff was necessary to achieve convergence in the three-body $\ntlo$ ${}^{4}\!P_{J}$, ${}^{2}\!P_{J}$, and ${}^{4}\!P_{J}\to{}^{2}\!P_{J}$ channels as they behave asymptotically like $q^{-0.545\cdots}$~\cite{Griesshammer:2005ga}, and therefore have slow convergence.  The asymptotic form of the ${}^{4}\!P_{J}$, ${}^{2}\!P_{J}$, and ${}^{4}\!P_{J}\to{}^{2}\!P_{J}$ channels will lead to the need for a three-body force at $\mathrm{N}^{3.5}\mathrm{LO}$~\cite{Griesshammer:2005ga}.  Some authors would advocate that order-by-order in the \EFT expansion the cutoff variation with respect to $\Lambda$ should grow smaller (for a discussion of this see Ref.~\cite{Griesshammer:2015osb}).  In such a power counting scheme the three-body force occurring at $\mathrm{N}^{3.5}\mathrm{LO}$ would be promoted to $\ntlo$.  Although this is a desirable property for a power counting it is not rigorously motivated in the three-body sector.\footnote{In the two-body sector it can be shown analytically using cutoff regularization that order-by-order the cutoff variation gets smaller.}  In future work we will calculate $\N{4}$ contributions, which include  the $\mathrm{N}^{3.5}\mathrm{LO}$ three-body force terms in the ${}^{4}\!P_{J}$, ${}^{2}\!P_{J}$, and ${}^{4}\!P_{J}\to{}^{2}\!P_{J}$ channels.  Unfortunately, these three-body forces must be fit to three-body data in the ${}^{4}\!P_{J}$, ${}^{2}\!P_{J}$, and ${}^{4}\!P_{J}\to{}^{2}\!P_{J}$ channels, which are expected to be important for obtaining an accurate description of $A_y$ because $P$-waves are the first angular momenta that cause splitting among $J$-values.  The three-body forces can be fit to $A_{y}$ at one energy and then predictions can be made for other energies.  At $\N{4}$ there will also be a new two-body $SD$-mixing term, a new energy dependent three-body force in the doublet $S$-wave channel, and a three-body $SD$-mixing term that can be fit to the asymptotic $D/S$-mixing ratio of the triton wavefunction.

In this \EFT calculation the doublet $S$-wave channel was only calculated to $\nnlo$.  The doublet $S$-wave channel has only one $J$-value and therefore no splitting for different $J$-values occurs.  As a result the doublet $S$-wave only contributes up to NLO in the numerator for all polarization observables. The denominator of all polarization observables is given by the unpolarized $nd$ scattering cross section.  In a strictly perturbative approach we would expand the denominator.  Since the first non-zero contribution to the numerator of polarization observables occurs at $\nnlo$ the doublet $S$-wave contribution from the expanded denominator would again only be needed to NLO for a $\ntlo$ calculation.  However, we find that the peak of $A_{y}$ depends on the minimum of the cross section that is only reproduced well at $\nnlo$.  Therefore we resum certain higher order contributions into the denominator and keep the cross section expanded perturbatively to $\nnlo$ in the denominator.  Future calculations will include the $\ntlo$ doublet $S$-wave channel and calculate the $\ntlo$ cross section.  This should not significantly change the results since good agreement is already observed with experimental data at $\nnlo$ for the cross section.  Calculation of the doublet $S$-wave channel to $\ntlo$ is likely complicated by the requirement of an additional Wigner-antisymmetric to Wigner-symmetric three-body force.  Griesshammer~\cite{Griesshammer:2005ga} claims that a Wigner-antisymmetric to Wigner-symmetric three-body force should occur at $\mathrm{N}^{5}\mathrm{LO}$ due to suppression from the Pauli principle, while Birse~\cite{Birse:2005pm} claims that it should occur at $\ntlo$ as the Pauli principle is already included in a naive asymptotic analysis.  Having fit $H_{\ntlo}$ to the doublet $S$-wave scattering length and $H_{2}^{(\ntlo)}$ to the triton binding energy we find the doublet $S$-wave $nd$ scattering amplitude is not properly renormalized, suggesting the need for a new three-body force as claimed by Birse.  Future work will address this new three-body force in order to have a complete $\ntlo$ calculation.  Also at $\ntlo$ and higher orders the divergences that must be renormalized become worse, leading to potential numerical issues, especially at higher cutoffs due to numerical fine tuning.

Finally, calculating polarization observables in $pd$ scattering is of interest due to the larger data set available for such interactions.  Such calculations are complicated due to Coulomb interactions, but have been performed in \EFT in the quartet and doublet $S$-wave channels in which Coulomb effects are treated ``perturbatively"\footnote{In Ref.~\cite{Vanasse:2014kxa} Coulomb is treated nonperturbatively in the two-body sector and in the three-body sector all one-photon exchange diagrams are resummed.  In Ref.~\cite{Konig:2015aka} Coulomb is treated strictly perturbatively in the two and three-body sector.}~\cite{Vanasse:2014kxa,Konig:2015aka}.  Higher partial waves will be needed but are in principle straightforward to include.  The main stumbling block to calculations in $pd$ scattering are three-body forces in the doublet $S$-wave channel.  It was shown in Ref.~\cite{Vanasse:2014kxa} that at NLO the same three-body forces could not be used to renormalize both $nd$ and $pd$ scattering.  This pattern likely persists to $\nnlo$ where a new energy dependent three-body force occurs.  In the $nd$ system the three-body forces are fit to the doublet $S$-wave $nd$ scattering length and the triton binding energy.  For $pd$ scattering fitting to the $\jjvHe$ binding energy is straightforward, but fitting the doublet $S$-wave $pd$ scattering length is complicated due to Coulomb effects.  Therefore an appropriate renormalization condition will need to be found for $pd$ scattering at $\nnlo$ in order to investigate the large $pd$ data set for polarization observables.

\appendix

\section{Appendix. Partial wave projection}

All of the diagrams used in these calculations need to be projected into the partial wave basis.  One approach is to construct all necessary projectors and use them to project all diagrams onto respective partial waves~\cite{Griesshammer:2004pe,Griesshammer:2011md}.  The advantage of this approach is that it makes projecting out diagrams very easy.  However, the downside is that a projector must be constructed for every channel of interest and only $S$- and $P$-wave projectors have been published to date.  Instead of the projector method we adopt a Racah algebra approach that, while computationally more intensive, gives all partial waves at once.  The contribution from a generic diagram is given by
\begin{equation}
 \left[\left(\boldsymbol{\mathcal{K}}^{jB}_{iA}(\vect{q},\vect{\boldsymbol{\ell}})\right)^{\beta b}_{\alpha a}\right]_{yx},
\end{equation}
where $i$ ($j$) is the initial (final) dibaryon spin polarization, $A$ ($B$) the initial (final) dibaryon isospin polarization, $\alpha$ ($\beta$) the initial (final) nucleon spin, and $a$ ($b$) the initial (final) nucleon isospin.  The subscripts $y$ and $x$ pick out a component of the c.c.~space matrix, with $x=1$ ($x=0$) corresponding to an initial spin-triplet (spin-singlet) dibaryon and $y=1$ ($y=0$) corresponding to a final spin-triplet (spin-singlet) dibaryon.  A generic contribution is projected onto a partial wave basis by
\begin{align}
\label{eq:projection}
&\left[\boldsymbol{\mathcal{K}}(q,\ell)^{J}_{L'S',LS}\right]_{yx}=\frac{1}{4\pi}\sum_{\chi}\CG{L}{m_{L}}{S}{m_{S}}{J}{M}\CG{L'}{m_{L'}}{S'}{m_{S'}}{J}{M}\CG{x}{i}{\frac{1}{2}}{\alpha}{S}{m_{S}}\CG{y}{j}{\frac{1}{2}}{\beta}{S'}{m_{S'}}\CG{1-x}{A}{\frac{1}{2}}{a}{\frac{1}{2}}{-\frac{1}{2}}\CG{1-y}{B}{\frac{1}{2}}{b}{\frac{1}{2}}{-\frac{1}{2}}\\\nonumber
&\hspace{2cm}\times\int\!\!d\Omega_{q}\int\!\!d\Omega_{\ell}\left[\left(\boldsymbol{\mathcal{K}}^{jB}_{iA}(\vect{q},\vect{\boldsymbol{\ell}})\right)^{\beta b}_{\alpha a}\right]_{yx}Y_{L}^{m_{L}}(\hat{\mathbf{q}})\left(Y_{L'}^{m_{L'}}(\hat{\boldsymbol{\ell}})\right)^{*},
\end{align}
where the first two Clebsch-Gordan coefficients couple orbital and spin angular momentum, the next two couple dibaryon and nucleon spin, and the final two couple nucleon and dibaryon isospin.  $\chi$ sums over all magnetic quantum numbers. In order to treat all elements of a c.c.~space matrix simultaneously we define the operators 
\begin{equation}
 S_{x}^{i}=\left\{\begin{array}{cc}
              1 & , \ \ x=0 \\
              \sigma_{i} & ,\ \ x=1
             \end{array}\right.\quad,\quad
T_{x}^{a}=\left\{\begin{array}{cc}
              1 & ,\ \ x=0 \\
              \tau_{a} & , \ \ x=1
             \end{array}\right. .
\end{equation}

Using the Wigner-Eckart theorem these operators when projected out give
\begin{equation}
\left\langle\frac{1}{2},m_{2}'\Big{|}S_{x}^{i}\Big{|}\frac{1}{2},m_{2}\right\rangle=\sqrt{\widehat{x}}\,\,\CG{\frac{1}{2}}{m_{2}}{x}{i}{\frac{1}{2}}{m_{2}'},
\end{equation}
and 

\begin{equation}
\left\langle\frac{1}{2},m_{2}'\Big{|}T_{x}^{a}\Big{|}\frac{1}{2},m_{2}\right\rangle=\sqrt{\widehat{x}}\,\,\CG{\frac{1}{2}}{m_{2}}{x}{a}{\frac{1}{2}}{m_{2}'}.
\end{equation}
As an example, a contribution from the $SD$-mixing term is given by
\begin{equation}
 \left[\left(\boldsymbol{\mathcal{K}}^{jB}_{iA}(\vect{q},\vect{\boldsymbol{\ell}})\right)^{\beta b}_{\alpha a}\right]^{(SD)}_{yx}=\frac{1}{a+\hat{\mathbf{q}}\cdot\hat{\boldsymbol{\ell}}}\left[\sigma_{m}{S_{y}^{j}}^{\dagger}{T_{1-y}^{B}}^{\dagger}\right]^{\beta b}_{\alpha a}\vect{q}_{i}\vect{q}_{-m}(-1)^{m}\delta_{1x}.
\end{equation}
Using Eq.~(\ref{eq:projection}) and the identity
\begin{align}
 &\frac{4\pi}{3}\ell^{2}\int\!\!d\Omega_{q}\int\!\!d\Omega_{\ell}\frac{1}{a+\hat{\mathbf{q}}\cdot\hat{\boldsymbol{\ell}}}Y_{L'}^{m_{L'}}(\hat{\boldsymbol{\ell}})^{*}Y_{L}^{m_{L}}(\hat{\mathbf{q}})Y_{1}^{m_{1}}(\hat{\boldsymbol{\ell}})Y_{1}^{m_{2}}(\hat{\boldsymbol{\ell}})=\\\nonumber
 &=4\pi \ell^{2}\sum_{L''}\sum_{m_{L''}}\sqrt{\frac{\widehat{L}}{\widehat{L'}}}\CG{1}{m_{1}}{1}{m_{2}}{L''}{m_{L''}}\CG{L}{m_{L}}{L''}{m_{L''}}{L'}{m_{L'}}\CG{1}{0}{1}{0}{L''}{0}\CG{L}{0}{L''}{0}{L'}{0}Q_{L}(a),
\end{align}
the projection of this contribution in the partial wave basis becomes
\begin{align}
 &\ell^{2}\sqrt{3\widehat{y}\widehat{(1-y)}}\,\sqrt{\frac{\widehat{L}}{\widehat{L'}}}\sum_{\chi}\CG{L}{m_{L}}{S}{m_{S}}{J}{M}\CG{L'}{m_{L'}}{S'}{m_{S'}}{J}{M}\CG{x}{i}{\frac{1}{2}}{\alpha}{S}{m_{S}}\CG{y}{j}{\frac{1}{2}}{\beta}{S'}{m_{S'}}\CG{1-x}{A}{\frac{1}{2}}{a}{\frac{1}{2}}{-\frac{1}{2}}\CG{1-y}{B}{\frac{1}{2}}{b}{\frac{1}{2}}{-\frac{1}{2}}\\\nonumber
&\CG{\frac{1}{2}}{m_{2}}{y}{-j}{\frac{1}{2}}{\widetilde{m_{2}}}\CG{\frac{1}{2}}{\widetilde{m_{2}}}{1}{m}{\frac{1}{2}}{m_{2}'}\CG{\frac{1}{2}}{a}{1-y}{-B}{\frac{1}{2}}{b}\CG{1}{i}{1}{-m}{L''}{m_{L''}}\CG{L}{m_{L}}{L''}{m_{L''}}{L'}{m_{L'}}\CG{1}{0}{1}{0}{L''}{0}\CG{L}{0}{L''}{0}{L}{0}Q_{L}(a)(-1)^{j+B+m}\delta_{1x}.
\end{align}
The sum over magnetic quantum numbers can then be simplified via Racah algebra yielding
\begin{align}
 &\left[\boldsymbol{\mathcal{K}}(q,\ell)^{J}_{L'S',LS}\right]^{(SD)}_{yx}=2\sqrt{3\widehat{y}\widehat{(1-y)}\widehat{S}\widehat{S'}\widehat{L}\widehat{L''}}(-1)^{\frac{1}{2}+x+y+L''+L+S+S'-J}\\\nonumber
 &\hspace{3cm}\times\SJ{y}{\frac{1}{2}}{\frac{1}{2}}{1}{S'}{\frac{1}{2}}\SJ{L''}{1}{x}{\frac{1}{2}}{S}{S'}\SJ{S'}{L''}{S}{L}{J}{L'}\CG{1}{0}{1}{0}{L''}{0}\CG{L}{0}{L''}{0}{L'}{0}\ell^{2}Q_{L}(a)\delta_{1x}.
\end{align}

\begin{acknowledgments}
We would like to thank D.R.~Phillips, M.R.~Schindler, and W. Tornow for discussions important for the completion of this work.  This material is based upon work supported by the U.S. Department of
Energy, Office of Science, Office of Nuclear Physics, under Award Number DE-FG02-05ER41368 and Award Number DE-FG02-93ER40756 (JV).
\end{acknowledgments}


\begin{thebibliography}{44}
\expandafter\ifx\csname natexlab\endcsname\relax\def\natexlab#1{#1}\fi
\expandafter\ifx\csname bibnamefont\endcsname\relax
  \def\bibnamefont#1{#1}\fi
\expandafter\ifx\csname bibfnamefont\endcsname\relax
  \def\bibfnamefont#1{#1}\fi
\expandafter\ifx\csname citenamefont\endcsname\relax
  \def\citenamefont#1{#1}\fi
\expandafter\ifx\csname url\endcsname\relax
  \def\url#1{\texttt{#1}}\fi
\expandafter\ifx\csname urlprefix\endcsname\relax\def\urlprefix{URL }\fi
\providecommand{\bibinfo}[2]{#2}
\providecommand{\eprint}[2][]{\url{#2}}

\bibitem[{\citenamefont{Entem et~al.}(2002)\citenamefont{Entem, Machleidt, and
  Witala}}]{Entem:2001tj}
\bibinfo{author}{\bibfnamefont{D.}~\bibnamefont{Entem}},
  \bibinfo{author}{\bibfnamefont{R.}~\bibnamefont{Machleidt}},
  \bibnamefont{and} \bibinfo{author}{\bibfnamefont{H.}~\bibnamefont{Witala}},
  \bibinfo{journal}{Phys. Rev. C} \textbf{\bibinfo{volume}{65}},
  \bibinfo{pages}{064005} (\bibinfo{year}{2002}), \eprint{nucl-th/0111033}.

\bibitem[{\citenamefont{Kievsky et~al.}(2013)\citenamefont{Kievsky, Viviani,
  and Marcucci}}]{Kievsky:2013yva}
\bibinfo{author}{\bibfnamefont{A.}~\bibnamefont{Kievsky}},
  \bibinfo{author}{\bibfnamefont{M.}~\bibnamefont{Viviani}}, \bibnamefont{and}
  \bibinfo{author}{\bibfnamefont{L.~E.} \bibnamefont{Marcucci}},
  \bibinfo{journal}{Few Body Syst.} \textbf{\bibinfo{volume}{54}},
  \bibinfo{pages}{2395} (\bibinfo{year}{2013}).

\bibitem[{\citenamefont{Binder et~al.}(2015)}]{Binder:2015mbz}
\bibinfo{author}{\bibfnamefont{S.}~\bibnamefont{Binder}} \bibnamefont{et~al.}
  (\bibinfo{year}{2015}), \eprint{1505.07218}.

\bibitem[{\citenamefont{Witała et~al.}(1989)\citenamefont{Witała, Glöckle,
  and Cornelius}}]{Witala1989446}
\bibinfo{author}{\bibfnamefont{H.}~\bibnamefont{Witała}},
  \bibinfo{author}{\bibfnamefont{W.}~\bibnamefont{Glöckle}}, \bibnamefont{and}
  \bibinfo{author}{\bibfnamefont{T.}~\bibnamefont{Cornelius}},
  \bibinfo{journal}{Nuclear Physics A} \textbf{\bibinfo{volume}{496}},
  \bibinfo{pages}{446 } (\bibinfo{year}{1989}), ISSN \bibinfo{issn}{0375-9474},
  \urlprefix\url{http://www.sciencedirect.com/science/article/pii/0375947489900705}.

\bibitem[{\citenamefont{Witała and Glöckle}(1991)}]{WITALA199148}
\bibinfo{author}{\bibfnamefont{H.}~\bibnamefont{Witała}} \bibnamefont{and}
  \bibinfo{author}{\bibfnamefont{W.}~\bibnamefont{Glöckle}},
  \bibinfo{journal}{Nuclear Physics A} \textbf{\bibinfo{volume}{528}},
  \bibinfo{pages}{48 } (\bibinfo{year}{1991}), ISSN \bibinfo{issn}{0375-9474},
  \urlprefix\url{http://www.sciencedirect.com/science/article/pii/0375947491904197}.

\bibitem[{\citenamefont{Tornow et~al.}(1991)\citenamefont{Tornow, Howell,
  Alohali, Chen, Felsher, Hanly, Walter, Weisel, Mertens, Slaus
  et~al.}}]{TORNOW1991273}
\bibinfo{author}{\bibfnamefont{W.}~\bibnamefont{Tornow}},
  \bibinfo{author}{\bibfnamefont{C.}~\bibnamefont{Howell}},
  \bibinfo{author}{\bibfnamefont{M.}~\bibnamefont{Alohali}},
  \bibinfo{author}{\bibfnamefont{Z.}~\bibnamefont{Chen}},
  \bibinfo{author}{\bibfnamefont{P.}~\bibnamefont{Felsher}},
  \bibinfo{author}{\bibfnamefont{J.}~\bibnamefont{Hanly}},
  \bibinfo{author}{\bibfnamefont{R.}~\bibnamefont{Walter}},
  \bibinfo{author}{\bibfnamefont{G.}~\bibnamefont{Weisel}},
  \bibinfo{author}{\bibfnamefont{G.}~\bibnamefont{Mertens}},
  \bibinfo{author}{\bibfnamefont{I.}~\bibnamefont{Slaus}},
  \bibnamefont{et~al.}, \bibinfo{journal}{Physics Letters B}
  \textbf{\bibinfo{volume}{257}}, \bibinfo{pages}{273 } (\bibinfo{year}{1991}),
  ISSN \bibinfo{issn}{0370-2693},
  \urlprefix\url{http://www.sciencedirect.com/science/article/pii/037026939191892Y}.

\bibitem[{\citenamefont{Huber and Friar}(1998)}]{Huber:1998hu}
\bibinfo{author}{\bibfnamefont{D.}~\bibnamefont{Huber}} \bibnamefont{and}
  \bibinfo{author}{\bibfnamefont{J.~L.} \bibnamefont{Friar}},
  \bibinfo{journal}{Phys. Rev. C} \textbf{\bibinfo{volume}{58}},
  \bibinfo{pages}{674} (\bibinfo{year}{1998}), \eprint{nucl-th/9803038}.

\bibitem[{\citenamefont{Beane et~al.}(2000)\citenamefont{Beane, Bedaque,
  Haxton, Phillips, and Savage}}]{Beane:2000fx}
\bibinfo{author}{\bibfnamefont{S.~R.} \bibnamefont{Beane}},
  \bibinfo{author}{\bibfnamefont{P.~F.} \bibnamefont{Bedaque}},
  \bibinfo{author}{\bibfnamefont{W.~C.} \bibnamefont{Haxton}},
  \bibinfo{author}{\bibfnamefont{D.~R.} \bibnamefont{Phillips}},
  \bibnamefont{and} \bibinfo{author}{\bibfnamefont{M.~J.} \bibnamefont{Savage}},
    \bibinfo{journal}{ In *Shifman, M. (ed.): At the frontier of
particle physics, vol. 1* pp. 133–269.}
  (\bibinfo{year}{2000}), \eprint{nucl-th/0008064}.

\bibitem[{\citenamefont{Chen et~al.}(1999)\citenamefont{Chen, Rupak, and
  Savage}}]{Chen:1999tn}
\bibinfo{author}{\bibfnamefont{J.-W.} \bibnamefont{Chen}},
  \bibinfo{author}{\bibfnamefont{G.}~\bibnamefont{Rupak}}, \bibnamefont{and}
  \bibinfo{author}{\bibfnamefont{M.~J.} \bibnamefont{Savage}},
  \bibinfo{journal}{Nucl. Phys. A} \textbf{\bibinfo{volume}{653}},
  \bibinfo{pages}{386} (\bibinfo{year}{1999}), \eprint{nucl-th/9902056}.

\bibitem[{\citenamefont{Kong and Ravndal}(2000)}]{Kong:1999sf}
\bibinfo{author}{\bibfnamefont{X.}~\bibnamefont{Kong}} \bibnamefont{and}
  \bibinfo{author}{\bibfnamefont{F.}~\bibnamefont{Ravndal}},
  \bibinfo{journal}{Nucl. Phys. A} \textbf{\bibinfo{volume}{665}},
  \bibinfo{pages}{137} (\bibinfo{year}{2000}), \eprint{hep-ph/9903523}.

\bibitem[{\citenamefont{Rupak}(2000)}]{Rupak:1999rk}
\bibinfo{author}{\bibfnamefont{G.}~\bibnamefont{Rupak}},
  \bibinfo{journal}{Nucl. Phys. A} \textbf{\bibinfo{volume}{678}},
  \bibinfo{pages}{405} (\bibinfo{year}{2000}), \eprint{nucl-th/9911018}.

\bibitem[{\citenamefont{Butler et~al.}(2001)\citenamefont{Butler, Chen, and
  Kong}}]{Butler:2000zp}
\bibinfo{author}{\bibfnamefont{M.}~\bibnamefont{Butler}},
  \bibinfo{author}{\bibfnamefont{J.-W.} \bibnamefont{Chen}}, \bibnamefont{and}
  \bibinfo{author}{\bibfnamefont{X.}~\bibnamefont{Kong}},
  \bibinfo{journal}{Phys. Rev. C} \textbf{\bibinfo{volume}{63}},
  \bibinfo{pages}{035501} (\bibinfo{year}{2001}), \eprint{nucl-th/0008032}.

\bibitem[{\citenamefont{Gabbiani et~al.}(2000)\citenamefont{Gabbiani, Bedaque,
  and Grie{\ss}hammer}}]{Gabbiani:1999yv}
\bibinfo{author}{\bibfnamefont{F.}~\bibnamefont{Gabbiani}},
  \bibinfo{author}{\bibfnamefont{P.~F.} \bibnamefont{Bedaque}},
  \bibnamefont{and} \bibinfo{author}{\bibfnamefont{H.~W.}
  \bibnamefont{Grie{\ss}hammer}}, \bibinfo{journal}{Nucl. Phys. A}
  \textbf{\bibinfo{volume}{675}}, \bibinfo{pages}{601} (\bibinfo{year}{2000}),
  \eprint{nucl-th/9911034}.

\bibitem[{\citenamefont{Grie{\ss}hammer}(2004)}]{Griesshammer:2004pe}
\bibinfo{author}{\bibfnamefont{H.~W.} \bibnamefont{Grie{\ss}hammer}},
  \bibinfo{journal}{Nucl. Phys. A} \textbf{\bibinfo{volume}{744}},
  \bibinfo{pages}{192} (\bibinfo{year}{2004}), \eprint{nucl-th/0404073}.

\bibitem[{\citenamefont{Vanasse}(2013)}]{Vanasse:2013sda}
\bibinfo{author}{\bibfnamefont{J.}~\bibnamefont{Vanasse}},
  \bibinfo{journal}{Phys. Rev. C} \textbf{\bibinfo{volume}{88}},
  \bibinfo{pages}{044001} (\bibinfo{year}{2013}), \eprint{1305.0283}.

\bibitem[{\citenamefont{Rupak and Kong}(2003)}]{Rupak:2001ci}
\bibinfo{author}{\bibfnamefont{G.}~\bibnamefont{Rupak}} \bibnamefont{and}
  \bibinfo{author}{\bibfnamefont{X.-w.} \bibnamefont{Kong}},
  \bibinfo{journal}{Nucl. Phys. A} \textbf{\bibinfo{volume}{717}},
  \bibinfo{pages}{73} (\bibinfo{year}{2003}), \eprint{nucl-th/0108059}.

\bibitem[{\citenamefont{K{\"o}nig and Hammer}(2011)}]{Konig:2011yq}
\bibinfo{author}{\bibfnamefont{S.}~\bibnamefont{K{\"o}nig}} \bibnamefont{and}
  \bibinfo{author}{\bibfnamefont{H.~W.} \bibnamefont{Hammer}},
  \bibinfo{journal}{Phys. Rev. C} \textbf{\bibinfo{volume}{83}},
  \bibinfo{pages}{064001} (\bibinfo{year}{2011}), \eprint{1101.5939}.

\bibitem[{\citenamefont{Vanasse et~al.}(2014)\citenamefont{Vanasse, Egolf,
  Kerin, König, and Springer}}]{Vanasse:2014kxa}
\bibinfo{author}{\bibfnamefont{J.}~\bibnamefont{Vanasse}},
  \bibinfo{author}{\bibfnamefont{D.~A.} \bibnamefont{Egolf}},
  \bibinfo{author}{\bibfnamefont{J.}~\bibnamefont{Kerin}},
  \bibinfo{author}{\bibfnamefont{S.}~\bibnamefont{König}}, \bibnamefont{and}
  \bibinfo{author}{\bibfnamefont{R.~P.} \bibnamefont{Springer}},
  \bibinfo{journal}{Phys. Rev.} \textbf{\bibinfo{volume}{C89}},
  \bibinfo{pages}{064003} (\bibinfo{year}{2014}), \eprint{1402.5441}.

\bibitem[{\citenamefont{König et~al.}(2015{\natexlab{a}})\citenamefont{König,
  Grießhammer, and Hammer}}]{Konig:2014ufa}
\bibinfo{author}{\bibfnamefont{S.}~\bibnamefont{König}},
  \bibinfo{author}{\bibfnamefont{H.~W.} \bibnamefont{Grießhammer}},
  \bibnamefont{and} \bibinfo{author}{\bibfnamefont{H.~W.}
  \bibnamefont{Hammer}}, \bibinfo{journal}{J. Phys.}
  \textbf{\bibinfo{volume}{G42}}, \bibinfo{pages}{045101}
  (\bibinfo{year}{2015}{\natexlab{a}}), \eprint{1405.7961}.

\bibitem[{\citenamefont{Bedaque et~al.}(2003)\citenamefont{Bedaque, Rupak,
  Grie{\ss}hammer, and Hammer}}]{Bedaque:2002yg}
\bibinfo{author}{\bibfnamefont{P.~F.} \bibnamefont{Bedaque}},
  \bibinfo{author}{\bibfnamefont{G.}~\bibnamefont{Rupak}},
  \bibinfo{author}{\bibfnamefont{H.~W.} \bibnamefont{Grie{\ss}hammer}},
  \bibnamefont{and} \bibinfo{author}{\bibfnamefont{H.-W.}
  \bibnamefont{Hammer}}, \bibinfo{journal}{Nucl. Phys. A}
  \textbf{\bibinfo{volume}{714}}, \bibinfo{pages}{589} (\bibinfo{year}{2003}),
  \eprint{nucl-th/0207034}.

\bibitem[{\citenamefont{Vanasse}()}]{Vanasse:2015}
\bibinfo{author}{\bibfnamefont{J.}~\bibnamefont{Vanasse}},
(\bibinfo{year}{2015}), \eprint{1512.03805}


\bibitem[{\citenamefont{Phillips et~al.}(2000)\citenamefont{Phillips, Rupak,
  and Savage}}]{Phillips:1999hh}
\bibinfo{author}{\bibfnamefont{D.~R.} \bibnamefont{Phillips}},
  \bibinfo{author}{\bibfnamefont{G.}~\bibnamefont{Rupak}}, \bibnamefont{and}
  \bibinfo{author}{\bibfnamefont{M.~J.} \bibnamefont{Savage}},
  \bibinfo{journal}{Phys. Lett. B} \textbf{\bibinfo{volume}{473}},
  \bibinfo{pages}{209} (\bibinfo{year}{2000}), \eprint{nucl-th/9908054}.

\bibitem[{\citenamefont{Kaplan et~al.}(1998{\natexlab{a}})\citenamefont{Kaplan,
  Savage, and Wise}}]{Kaplan:1998tg}
\bibinfo{author}{\bibfnamefont{D.~B.} \bibnamefont{Kaplan}},
  \bibinfo{author}{\bibfnamefont{M.~J.} \bibnamefont{Savage}},
  \bibnamefont{and} \bibinfo{author}{\bibfnamefont{M.~B.} \bibnamefont{Wise}},
  \bibinfo{journal}{Phys. Lett. B} \textbf{\bibinfo{volume}{424}},
  \bibinfo{pages}{390} (\bibinfo{year}{1998}{\natexlab{a}}),
  \eprint{nucl-th/9801034}.

\bibitem[{\citenamefont{Kaplan et~al.}(1998{\natexlab{b}})\citenamefont{Kaplan,
  Savage, and Wise}}]{Kaplan:1998we}
\bibinfo{author}{\bibfnamefont{D.~B.} \bibnamefont{Kaplan}},
  \bibinfo{author}{\bibfnamefont{M.~J.} \bibnamefont{Savage}},
  \bibnamefont{and} \bibinfo{author}{\bibfnamefont{M.~B.} \bibnamefont{Wise}},
  \bibinfo{journal}{Nucl. Phys. B} \textbf{\bibinfo{volume}{534}},
  \bibinfo{pages}{329} (\bibinfo{year}{1998}{\natexlab{b}}),
  \eprint{nucl-th/9802075}.

\bibitem[{\citenamefont{Stoks et~al.}(1994)\citenamefont{Stoks, Klomp,
  Terheggen, and de~Swart}}]{Stoks:1994wp}
\bibinfo{author}{\bibfnamefont{V.}~\bibnamefont{Stoks}},
  \bibinfo{author}{\bibfnamefont{R.}~\bibnamefont{Klomp}},
  \bibinfo{author}{\bibfnamefont{C.}~\bibnamefont{Terheggen}},
  \bibnamefont{and} \bibinfo{author}{\bibfnamefont{J.}~\bibnamefont{de~Swart}},
  \bibinfo{journal}{Phys. Rev. C} \textbf{\bibinfo{volume}{49}},
  \bibinfo{pages}{2950} (\bibinfo{year}{1994}), \eprint{nucl-th/9406039}.

\bibitem[{\citenamefont{Chen and Savage}(1999)}]{Chen:1999bg}
\bibinfo{author}{\bibfnamefont{J.-W.} \bibnamefont{Chen}} \bibnamefont{and}
  \bibinfo{author}{\bibfnamefont{M.~J.} \bibnamefont{Savage}},
  \bibinfo{journal}{Phys. Rev. C} \textbf{\bibinfo{volume}{60}},
  \bibinfo{pages}{065205} (\bibinfo{year}{1999}), \eprint{nucl-th/9907042}.

\bibitem[{\citenamefont{Grie{\ss}hammer}(2005)}]{Griesshammer:2005ga}
\bibinfo{author}{\bibfnamefont{H.~W.} \bibnamefont{Grie{\ss}hammer}},
  \bibinfo{journal}{Nucl. Phys. A} \textbf{\bibinfo{volume}{760}},
  \bibinfo{pages}{110} (\bibinfo{year}{2005}), \eprint{nucl-th/0502039}.

\bibitem[{\citenamefont{Birse}(2006)}]{Birse:2005pm}
\bibinfo{author}{\bibfnamefont{M.~C.} \bibnamefont{Birse}},
  \bibinfo{journal}{J. Phys.} \textbf{\bibinfo{volume}{A39}},
  \bibinfo{pages}{L49} (\bibinfo{year}{2006}), \eprint{nucl-th/0509031}.

\bibitem[{\citenamefont{Hetherington and Schick}(1965)}]{Hetherington:1965zza}
\bibinfo{author}{\bibfnamefont{J.~H.} \bibnamefont{Hetherington}}
  \bibnamefont{and} \bibinfo{author}{\bibfnamefont{L.~H.}
  \bibnamefont{Schick}}, \bibinfo{journal}{Phys. Rev.}
  \textbf{\bibinfo{volume}{137}}, \bibinfo{pages}{B935} (\bibinfo{year}{1965}).

\bibitem[{\citenamefont{Aaron and Amado}(1966)}]{Aaron:1966zz}
\bibinfo{author}{\bibfnamefont{R.}~\bibnamefont{Aaron}} \bibnamefont{and}
  \bibinfo{author}{\bibfnamefont{R.~D.} \bibnamefont{Amado}},
  \bibinfo{journal}{Phys. Rev.} \textbf{\bibinfo{volume}{150}},
  \bibinfo{pages}{857} (\bibinfo{year}{1966}).

\bibitem[{\citenamefont{Schmid and Ziegelmann}(1974)}]{Ziegelmann}
\bibinfo{author}{\bibfnamefont{E.}~\bibnamefont{Schmid}} \bibnamefont{and}
  \bibinfo{author}{\bibfnamefont{H.}~\bibnamefont{Ziegelmann}},
  \emph{\bibinfo{title}{The Qauntum Mechanical Three-Body Problem, Vieweg Tract
  in Pure and Applied Physics Vol. 2}} (\bibinfo{publisher}{Pergamon Press},
  \bibinfo{year}{1974}).

\bibitem[{\citenamefont{Darden et~al.}(1971)\citenamefont{Darden, Barschall,
  and Haeberli}}]{darden1971polarization}
\bibinfo{author}{\bibfnamefont{S.}~\bibnamefont{Darden}},
  \bibinfo{author}{\bibfnamefont{H.}~\bibnamefont{Barschall}},
  \bibnamefont{and} \bibinfo{author}{\bibfnamefont{W.}~\bibnamefont{Haeberli}},
   \emph{\bibinfo{title}{Polarization Phenomena in Nuclear Reactions}}
  \bibinfo{journal}{University of Wisconsin press, Madison}
  p.~\bibinfo{pages}{39} (\bibinfo{year}{1971}).

\bibitem[{\citenamefont{Moeini~Arani and Bayegan}(2013)}]{Arani:2011if}
\bibinfo{author}{\bibfnamefont{M.}~\bibnamefont{Moeini~Arani}}
  \bibnamefont{and} \bibinfo{author}{\bibfnamefont{S.}~\bibnamefont{Bayegan}},
  \bibinfo{journal}{Eur. Phys. J.} \textbf{\bibinfo{volume}{A49}},
  \bibinfo{pages}{117} (\bibinfo{year}{2013}), \eprint{1111.0391}.

\bibitem[{\citenamefont{Ohlsen}(1972)}]{Ohlsen:1972zz}
\bibinfo{author}{\bibfnamefont{G.~G.} \bibnamefont{Ohlsen}},
  \bibinfo{journal}{Rept. Prog. Phys.} \textbf{\bibinfo{volume}{35}},
  \bibinfo{pages}{717} (\bibinfo{year}{1972}).

\bibitem[{\citenamefont{Gl{\"o}ckle}(2012)}]{glockle2012quantum}
\bibinfo{author}{\bibfnamefont{W.}~\bibnamefont{Gl{\"o}ckle}},
  \emph{\bibinfo{title}{The quantum mechanical few-body problem}}
  (\bibinfo{publisher}{Springer Science \& Business Media},
  \bibinfo{year}{2012}).

\bibitem[{\citenamefont{Fukukawa and Fujiwara}(2011)}]{Fukukawa:2011wu}
\bibinfo{author}{\bibfnamefont{K.}~\bibnamefont{Fukukawa}} \bibnamefont{and}
  \bibinfo{author}{\bibfnamefont{Y.}~\bibnamefont{Fujiwara}},
  \bibinfo{journal}{Prog. Theor. Phys.} \textbf{\bibinfo{volume}{125}},
  \bibinfo{pages}{729} (\bibinfo{year}{2011}), \eprint{1101.2977}.

\bibitem[{\citenamefont{Schwarz et~al.}(1983)\citenamefont{Schwarz, Klages,
  Doll, Haesner, Wilczynski, Zeitnitz, and Kecskemeti}}]{Schwarz19831}
\bibinfo{author}{\bibfnamefont{P.}~\bibnamefont{Schwarz}},
  \bibinfo{author}{\bibfnamefont{H.}~\bibnamefont{Klages}},
  \bibinfo{author}{\bibfnamefont{P.}~\bibnamefont{Doll}},
  \bibinfo{author}{\bibfnamefont{B.}~\bibnamefont{Haesner}},
  \bibinfo{author}{\bibfnamefont{J.}~\bibnamefont{Wilczynski}},
  \bibinfo{author}{\bibfnamefont{B.}~\bibnamefont{Zeitnitz}}, \bibnamefont{and}
  \bibinfo{author}{\bibfnamefont{J.}~\bibnamefont{Kecskemeti}},
  \bibinfo{journal}{Nuclear Physics A} \textbf{\bibinfo{volume}{398}},
  \bibinfo{pages}{1 } (\bibinfo{year}{1983}), ISSN \bibinfo{issn}{0375-9474},
  \urlprefix\url{http://www.sciencedirect.com/science/article/pii/0375947483906450}.

\bibitem[{\citenamefont{Kievsky et~al.}(1996)\citenamefont{Kievsky, Rosati,
  Tornow, and Viviani}}]{Kievsky:1996ca}
\bibinfo{author}{\bibfnamefont{A.}~\bibnamefont{Kievsky}},
  \bibinfo{author}{\bibfnamefont{S.}~\bibnamefont{Rosati}},
  \bibinfo{author}{\bibfnamefont{W.}~\bibnamefont{Tornow}}, \bibnamefont{and}
  \bibinfo{author}{\bibfnamefont{M.}~\bibnamefont{Viviani}},
  \bibinfo{journal}{Nucl. Phys. A} \textbf{\bibinfo{volume}{607}},
  \bibinfo{pages}{402} (\bibinfo{year}{1996}).

\bibitem[{\citenamefont{Neidel et~al.}(2003)\citenamefont{Neidel, Tornow,
  Trotter, Howell, Crowell, Macri, Walter, Weisel, Esterline, Wita{\l}a
  et~al.}}]{neidel2003new}
\bibinfo{author}{\bibfnamefont{E.}~\bibnamefont{Neidel}},
  \bibinfo{author}{\bibfnamefont{W.}~\bibnamefont{Tornow}},
  \bibinfo{author}{\bibfnamefont{D.~G.} \bibnamefont{Trotter}},
  \bibinfo{author}{\bibfnamefont{C.}~\bibnamefont{Howell}},
  \bibinfo{author}{\bibfnamefont{A.}~\bibnamefont{Crowell}},
  \bibinfo{author}{\bibfnamefont{R.}~\bibnamefont{Macri}},
  \bibinfo{author}{\bibfnamefont{R.}~\bibnamefont{Walter}},
  \bibinfo{author}{\bibfnamefont{G.}~\bibnamefont{Weisel}},
  \bibinfo{author}{\bibfnamefont{J.}~\bibnamefont{Esterline}},
  \bibinfo{author}{\bibfnamefont{H.}~\bibnamefont{Wita{\l}a}},
  \bibnamefont{et~al.}, \bibinfo{journal}{Physics Letters B}
  \textbf{\bibinfo{volume}{552}}, \bibinfo{pages}{29} (\bibinfo{year}{2003}).

\bibitem[{\citenamefont{McAninch et~al.}(1994)\citenamefont{McAninch, Lamm, and
  Haeberli}}]{McAninch:1994zz}
\bibinfo{author}{\bibfnamefont{J.~E.} \bibnamefont{McAninch}},
  \bibinfo{author}{\bibfnamefont{L.~O.} \bibnamefont{Lamm}}, \bibnamefont{and}
  \bibinfo{author}{\bibfnamefont{W.}~\bibnamefont{Haeberli}},
  \bibinfo{journal}{Phys. Rev.} \textbf{\bibinfo{volume}{C50}},
  \bibinfo{pages}{589} (\bibinfo{year}{1994}).

\bibitem[{\citenamefont{Shimizu et~al.}(1995)\citenamefont{Shimizu, Sagara,
  Nakamura, Maeda, Miwa, Nishimori, Ueno, Nakashima, and
  Morinobu}}]{Shimizu:1995zz}
\bibinfo{author}{\bibfnamefont{S.}~\bibnamefont{Shimizu}},
  \bibinfo{author}{\bibfnamefont{K.}~\bibnamefont{Sagara}},
  \bibinfo{author}{\bibfnamefont{H.}~\bibnamefont{Nakamura}},
  \bibinfo{author}{\bibfnamefont{K.}~\bibnamefont{Maeda}},
  \bibinfo{author}{\bibfnamefont{T.}~\bibnamefont{Miwa}},
  \bibinfo{author}{\bibfnamefont{N.}~\bibnamefont{Nishimori}},
  \bibinfo{author}{\bibfnamefont{S.}~\bibnamefont{Ueno}},
  \bibinfo{author}{\bibfnamefont{T.}~\bibnamefont{Nakashima}},
  \bibnamefont{and} \bibinfo{author}{\bibfnamefont{S.}~\bibnamefont{Morinobu}},
  \bibinfo{journal}{Phys. Rev.} \textbf{\bibinfo{volume}{C52}},
  \bibinfo{pages}{1193} (\bibinfo{year}{1995}).

\bibitem[{\citenamefont{Griesshammer}(2015)}]{Griesshammer:2015osb}
\bibinfo{author}{\bibfnamefont{H.~W.} \bibnamefont{Griesshammer}}, in
  \emph{\bibinfo{booktitle}{{8th International Workshop on Chiral Dynamics (CD
  2015) Pisa, Italy, June 29-July 3, 2015}}} (\bibinfo{year}{2015}),
  \eprint{1511.00490},
  \urlprefix\url{https://inspirehep.net/record/1402387/files/arXiv:1511.00490.pdf}.

\bibitem[{\citenamefont{König et~al.}(2015{\natexlab{b}})\citenamefont{König,
  Grießhammer, Hammer, and van Kolck}}]{Konig:2015aka}
\bibinfo{author}{\bibfnamefont{S.}~\bibnamefont{König}},
  \bibinfo{author}{\bibfnamefont{H.~W.} \bibnamefont{Grießhammer}},
  \bibinfo{author}{\bibfnamefont{H.~W.} \bibnamefont{Hammer}},
  \bibnamefont{and} \bibinfo{author}{\bibfnamefont{U.}~\bibnamefont{van Kolck}}
  (\bibinfo{year}{2015}{\natexlab{b}}), \eprint{1508.05085}.

\bibitem[{\citenamefont{Grie{\ss}hammer
  et~al.}(2012)\citenamefont{Grie{\ss}hammer, Schindler, and
  Springer}}]{Griesshammer:2011md}
\bibinfo{author}{\bibfnamefont{H.~W.} \bibnamefont{Grie{\ss}hammer}},
  \bibinfo{author}{\bibfnamefont{M.~R.} \bibnamefont{Schindler}},
  \bibnamefont{and} \bibinfo{author}{\bibfnamefont{R.~P.}
  \bibnamefont{Springer}}, \bibinfo{journal}{Eur. Phys. J. A}
  \textbf{\bibinfo{volume}{48}}, \bibinfo{pages}{7} (\bibinfo{year}{2012}),
  \eprint{1109.5667}.

\end{thebibliography}

\end{document}